\documentclass[preprint]{emulateapj}
\usepackage{amsmath,amssymb,graphicx,longtable,booktabs,lscape}

\graphicspath{}

\newcommand{\kms} {$\mbox{km s}^{-1}$}

\newcommand{\Msun} {$\mbox{M}_{\sun}$}

\newcommand{\kpc}{\,{\rm kpc}}
\newcommand{\mpc}{\,{\rm Mpc}}

\newcommand{\hbeta}{H$\beta$}

\newcommand{\mgb}{Mg$\,b$}

\newcommand{\atlas}{{ATLAS$^{\rm 3D}$}}
\newcommand{\lr}{{$\lambda_R$}}
\newcommand{\gmosifu}{GMOS-IFU}

\shorttitle{NGVS. XII. Stellar Populations and Kinematics of Compact, Low-Mass Early-Type Galaxies}
\shortauthors{Gu\'{e}rou et al.}

\begin{document}

\title{The Next Generation Virgo Cluster Survey. XII. Stellar Populations and Kinematics of Compact, Low-Mass Early-Type Galaxies from Gemini \gmosifu\ Spectroscopy}
\author{Adrien Gu\'{e}rou\altaffilmark{1,2,3,*}, Eric Emsellem\altaffilmark{3,4}, Richard M. McDermid\altaffilmark{5,6}, Patrick C\^{o}t\'{e}\altaffilmark{7}, Laura Ferrarese\altaffilmark{7}, John P. Blakeslee,\altaffilmark{7} Patrick R. Durrell\altaffilmark{8}, Lauren A. MacArthur\altaffilmark{7,9} , Eric W. Peng\altaffilmark{10,11}, Jean-Charles Cuillandre\altaffilmark{12}, Stephen Gwyn\altaffilmark{13}}

\altaffiltext{1}{IRAP, Institut de Recherche en Astrophysique et Plané\'{e}ologie, CNRS, 14, avenue Edouard Belin, F-31400 Toulouse, France}
\altaffiltext{2}{IRAP, Universit\'{e} de Toulouse, UPS-OMP, Toulouse, France}
\altaffiltext{3}{European Southern Observatory, Karl-Schwarzschild-Str. 2, D-85748 Garching, Germany}
\altaffiltext{4}{Universit\'{e} de Lyon 1, CRAL, Observatoire de Lyon, 9 av. Charles Andr\'{e}, F-69230 Saint-Genis Laval; CNRS, UMR 5574; ENS de Lyon, France}
\altaffiltext{5}{Department of Physics and Astronomy, Macquarie University, Sydney NSW 2109, Australia}
\altaffiltext{6}{Australian Gemini Office, Australian Astronomical Observatory, PO Box 915, Sydney NSW 1670, Australia}
\altaffiltext{7}{National Research Council of Canada, Herzberg Astronomy and Astrophysics, 5071 W. Saanich Road,Victoria, BC V9E 2E7, Canada}
\altaffiltext{8}{Department of Physics and Astronomy, Youngstown State University, Youngstown, OH 44555, USA}
\altaffiltext{9}{Department of Astrophysical Sciences, Princeton University, Princeton, NJ 08544, USA}
\altaffiltext{10}{Department of Astronomy, Peking University, Beijing 100871, China}
\altaffiltext{11}{Kavli Institute for Astronomy and Astrophysics, Peking University, Beijing 100871, China}
\altaffiltext{12}{CEA/IRFU/SAP, Laboratoire AIM Paris-Saclay, CNRS/INSU, Universit\'{e} Paris Diderot, Observatoire de Paris, PSL Research University, F-91191 Gif-sur-Yvette Cedex, France}
\altaffiltext{13}{Canadian Astronomy Data Centre, 5071 West Saanich Rd, Victoria BC, V9E 2E7, Canada} 
\altaffiltext{*}{aguerou@eso.org}

\begin{abstract}
We present Gemini \gmosifu\ data of eight compact low-mass early-type galaxies (ETGs) in the Virgo cluster. We analyse their stellar kinematics, stellar population, and present two-dimensional maps of these properties covering the central $5\arcsec\times7\arcsec$ region. We find a large variety of kinematics: from non- to highly-rotating objects, often associated with underlying disky isophotes revealed by deep images from the Next Generation Virgo Cluster Survey. In half of our objects, we find a centrally-concentrated younger and more metal-rich stellar population. We analyze the specific stellar angular momentum through the \lr\ parameter and find six fast-rotators and two slow-rotators, one having a thin counter-rotating disk. We compare the local galaxy density and stellar populations of our objects with those of 39 more extended low-mass Virgo ETGs from the SMAKCED survey and 260 massive ($M>10^{10}$\Msun) ETGs from the \atlas\ sample. The compact low-mass ETGs in our sample are located in high density regions, often close to a massive galaxy and have, on average, older and more metal-rich stellar populations than less compact low-mass galaxies. We find that the stellar population parameters follow lines of constant velocity dispersion in the mass-size plane, smoothly extending the comparable trends found for massive ETGs. Our study supports a scenario where low-mass compact ETGs have experienced long-lived interactions with their environment, including ram-pressure stripping and gravitational tidal forces, that may be responsible for their compact nature.
\end{abstract}

\section{Introduction}

Bright galaxies are known to obey well-defined sets of scaling relations linking their structural, kinematic and stellar population properties. During the past decade, a new generation of observational surveys based on optical integral-field unit (IFU) spectroscopy --- and supplemented with spectroscopic and imaging data covering a wide wavelength range --- has greatly improved our understanding of these high-luminosity  galaxies. Notable examples of such surveys include SAURON \citep{TimdeZeeuw2002}, \atlas\ \citep{Cappellari2011a} and CALIFA \citep{Sanchez2012}. 

The \atlas\ survey, which  carefully examined the structure, kinematics, gas-content and stellar populations of 260 early-type galaxies (ETGs) using the SAURON instrument on the 4.2m William Herschel Telescope, has provided many new insights into the formation and evolution of galaxies belonging to this broad class. Employing a $K_s$-band magnitude selection, complete down of $M_K < -21.5$ for morphologically ETGs within 42~Mpc, \atlas\ thoroughly sampled the ETG population down to a stellar mass of $M_{\ast} \simeq 6\times10^9~M_{\odot}$. As it happens, this limit coincides roughly with the well-known inflection point in the  mass-size and mass-surface brightness relations that are sometimes taken to separate the ETGs into high- and low-mass classes (see, e.g., Figure 16 of \citealt{Ferrarese2012a}). The low-mass ETGs themselves appear to subdivide into separate branches: diffuse and compact (see, e.g., Fig.~2 of \citealt{Dabringhausen2008}) suggesting different origins. Based on their simulations, diffuse low-mass ETGs are thought be formed mainly through tidal interactions in a weak field, whereas compact low-mass galaxies would be the low-mass counter parts of massive ETGs. It is difficult from imaging alone to assess the extent to which low-mass ETGs have properties similar to high-mass ETGs, and whether or not they have experienced similar formation and evolutionary processes (e.g., hierarchical merging, violent relaxation, cold gas accretion, tidal stirring and/or stripping, ram pressure stripping, etc). Whereas strong evidence, both from observations~\citep{Huxor2011a, Seth2014a} and simulations~\citep{Okazaki2000, Metz2007} exists about the importance of tidal stripping in the formation of low-mass ETGs, the respective roles of mergers and gas accretion is not clear yet, while the latter are prominent for more massive ETGs. In clusters, ram-pressure stripping may also be an efficient driver for the transformation of late-type star-forming galaxies into low-mass ETGs \citep{Vollmer2007, Boselli2009, Fumagalli2014}.

There have been several recent attempts to provide an initial characterization of the kinematics and stellar populations of low-mass ETGs using long-slit or IFU spectroscopy (e.g., \citealt{Michielsen2008, Chilingarian2009a, Koleva2011, Paudel2011, Rys2014, Toloba2014cLR}). These studies have focused almost entirely on low-mass ETGs that have generally been classified as dE or dS0 types (i.e., comparatively diffuse objects of both the nucleated and non-nucleated varieties). These are obvious first targets as such morphological classes make up the great majority of low-mass ETGs. 

A general finding from these studies is that low-mass ETGs show a surprising variety of kinematic structures: features that are remarkably similar to what has been observed in the more massive ETGs. The low-mass ETGs are also found to exhibit a wide range in age and metallicity. Thus, in terms of their stellar content, kinematics and structural parameters, they seem to bridge the gap between the faintest ETGs (such as the ``dwarf spheroidals" found in the Local Group) and the massive ETGs studied by \atlas\,. Still, the number of low-mass ETGs that have been examined using high-quality spectroscopy remains small, and existing samples include almost none of the compact (i.e., high-surface brightness) ETGs that appear to form a second branch in the photometric scaling relations. Although such systems are known to be very rare (see, e.g., \citealt{Bender1992}), a full understanding of ETGs will remain elusive until the nature of these objects can be understood with the benefit of high-quality (and preferably two-dimensional) spectroscopy.

In this paper, we carry out a spectroscopic survey of low-mass, compact ETGs in the Virgo cluster using IFU spectroscopy acquired with the 8.1m Gemini North Telescope. Our target galaxies have been selected from both the pioneering photographic survey of \citet{Binggeli1985} and from our own Next Generation Virgo Cluster Survey (NGVS; \citealt{Ferrarese2012a}). The NGVS is a large ($\simeq$ 900 hours) program carried out with the 3.6m Canada-France-Hawaii Telescope (CFHT) between 2008 and 2013. The survey  used the MegaCam instrument to perform panoramic imaging of the Virgo cluster, from its core to virial radius, in the $u^*giz$ filters (a total area of $\approx$ 100 deg$^{2}$). For a subset of the survey area, $r$- and $K$-band imaging is also available (see, e.g., \citealt{Munoz2014}). The point-source completeness limit for the survey is roughly $g \approx 25.9$~mag (10$\sigma$), with a 
corresponding surface brightness limit of $\mu_g \approx$ 29 mag~arcsec$^{-2}$ (2$\sigma$ above sky level). The NGVS image quality is also excellent, with a median $i$-band seeing of FWHM $\approx 0\farcs54$. A key deliverable from the NGVS will be an updated catalog of confirmed and probable Virgo cluster members, including a complete census of the low-mass compact ETGs that are the focus of this paper.

Other papers in the NGVS series related to the general topics considered here include studies of the globular cluster populations in Virgo \citep{Durrell2014}, the dynamical properties of star clusters, ultra-compact dwarfs (UCDs) and galaxies in the cluster core (\citealt{Zhu2014a}; Zhang et~al.~2015 (submitted)), a census of UCDs in Virgo's three main subclusters and an examination of tidal stripping as a possible origin for at least some UCDs (Liu~et~al.~in preparation), and investigations into ongoing tidal interactions of galaxies within the cluster environment \citep{ArrigoniBattaia2012, Paudel2013}. 

This paper is organized as follows. We describe our sample selection and data reduction procedures in \S\ref{sec:data}, and discuss our kinematical and stellar population analysis in \S\ref{sec:analysis}. Our results are presented in \S\ref{sec:results} and discussed in \S\ref{sec:discussion}, wherein we examine possible formation scenarios for low-mass ETGs in cluster environments. We summarize our findings in \S\ref{sec:summary}.

\section{Data}
\label{sec:data}

\subsection{Sample selection}
\label{sec:sample}

\begin{deluxetable*}{clccccccccc}
 \tablewidth{0pt}
 \tablecaption{Observed and Derived Properties of GMOS Target Galaxies
 \label{tab:data}}
 \tablehead{
  \colhead{Galaxy} &
  \colhead{Other ID} &	
  \colhead{RA(J2000)} &
  \colhead{DEC(J2000)} &
  \colhead{Distance} &
  \colhead{d(M87)} &
  \colhead{d(M49)} &
  \colhead{m$_{i}$} &
  \colhead{${\mu_{i}}$} &
  \colhead{R$_{e}$} &
  \colhead{$\epsilon$} \\
  \colhead{} &
  \colhead{} &
  \colhead{(hh:mm:ss.ss)} &
  \colhead{(dd:mm:ss.s)} &
  \colhead{(Mpc)} &
  \colhead{(Mpc)} &
  \colhead{(Mpc)} &
  \colhead{(mag)} &
  \colhead{(mag~arcsec$^{-2}$)} &
  \colhead{(arcsec)} &
  \colhead{} \\
  \colhead{(1)} &
  \colhead{(2)} &
  \colhead{(3)} &
  \colhead{(4)} &
  \colhead{(5)} &
  \colhead{(6)} &
  \colhead{(7)} &
  \colhead{(8)} &
  \colhead{(9)} &
  \colhead{(10)} &
  \colhead{(11)} }
  \startdata
	VCC\,0032 & IC\,0767   & 12:11:02.73 & +12:06:14.4 & 16.5$^{4}$ & 1.394 & 1.777 & 13.12 & 18.85 & 5.60 & 0.18\\
	VCC\,1178 & NGC\,4464  & 12:29:21.29 & +08:09:23.8 & 15.8$^{1}$ & 1.172 & 0.052 & 11.90 & 17.91 & 6.34 & 0.30\\
	VCC\,1192 & NGC\,4467  & 12:29:30.25 & +07:59:34.3 & 16.3$^{2}$ & 1.255 & 0.019 & 13.54 & 19.10 & 5.17 & 0.35\\
	VCC\,1199 & -          & 12:29:34.99 & +08:03:28.8 & 16.3$^{2}$ & 1.236 & 0.021 & 14.84 & 18.43 & 2.09 & 0.20\\
	VCC\,1297 & NGC\,4486b & 12:30:31.97 & +12:29:24.6 & 16.3$^{1}$ & 0.035 & 1.278 & 12.62 & 16.72 & 2.64 & 0.09\\
	VCC\,1440 & IC\,0798   & 12:32:33.41 & +15:24:55.4 & 16.1$^{3}$ & 0.858 & 2.092 & 13.58 & 20.10 & 8.06 & 0.01\\
	VCC\,1475 & NGC\,4515  & 12:33:04.97 & +16:15:55.9 & 16.7$^{3}$ & 1.140 & 2.420 & 12.01 & 18.81 & 9.13 & 0.35\\
	VCC\,1627 & -          & 12:35:37.25 & +12:22:55.4 & 15.6$^{1}$ & 0.319 & 1.255 & 13.89 & 18.74 & 3.71 & 0.10
  \enddata
\tablecomments{(1) Galaxy name from the catalogue of \citet{Binggeli1985}, (2) Alternative galaxy name in the NGC or IC catalogs, (3)-(4) J2000 coordinates (NED), (5) Redshift-independent distance when available (\citealt[$^{1}$][]{Mei2007}, \citealt[$^2$][]{Jordan2007}, \citealt[$^3$][]{Blakeslee2009}) or mean Virgo distance$^{4}$, (6)-(7) Projected distances in Mpc from M\,87 (NGC\,4486) and M\,49 (NGC\,4472), respectively, (8)-(9) Apparent magnitude and surface brightness measured from NGVS \textit{i}-band images, (10) Effective radius of the galaxy measured from the curve of growth of NGVS \textit{i}-band images, (11) Ellipticity derived as a moment of the surface brightness of the NGVS \textit{i}-band images, at $R_{e}/2$, except for two targets, VCC\,1440 and VCC\,1475, that we only cover up to $R_{e}/3$ with our \gmosifu\, observations.}
\end{deluxetable*}

The Virgo cluster is the galaxy cluster nearest to the Milky Way, containing nearly 2000 cataloged members  \citep{Binggeli1985}. The majority of these are classified as low-luminosity ETGs, thus making them a very significant morphological group by number, and an important sample for understanding galaxy evolution in dense environments. We selected our targets from \cite{Binggeli1985} as well as from the NGVS.

We first identified Virgo ETGs \citep{Binggeli1985} with $M_{B}<-18.0$, and then selected compact, high surface brightness objects that were objectively defined as lying more than 1$\sigma$ above the ridge-line in the magnitude-surface brightness diagram shown in Fig.~1 of \cite{Ferrarese2012a}. This approach is motivated by the fact that more massive galaxies ($M_{\ast}\gtrsim$10$^9$~M$_\odot$) generally follow this relation within a 1$\sigma$ limit, whereas some low-mass ETGs lie well above the relation, a region where tidally-stripped galaxies may reside (e.g., \citealt{King1962, Faber1973, Keenan1975}). Recently, \cite{Toloba2014bStellPop, Toloba2014cLR}, as well as \cite{Rys2013}, have suggested that {\it most} low-mass ETGs (i.e., systems with lower surface brightness than the objects targeted in this study) are the remains of galaxies that have been strongly transformed by ram-pressure stripping and gravitational harassment after falling into the cluster. \cite{Rys2014} also found that their stellar mass has contracted relative to their progenitors (assumed to be late-type). Therefore, our CFHT/Gemini survey --- which targets some of the most compact (i.e.\ highest surface brightness) Virgo ETGs with $i \gtrsim 12$ --- should provide new insights into the formation of these extreme objects.

We obtained Gemini data for eight of the nineteen galaxies that comprised our parent low-mass, compact ETG sample in the Virgo cluster, with bad weather conditions precluding observations of the remaining galaxies. Despite the fact that our actual sample is not complete, it should nevertheless be representative and allows us to probe a size-mass regime that has not yet been surveyed with the benefit of Integral Field Spectroscopy (IFS) data. The galaxies in our sample have surface brightnesses in the range ${17 \lesssim \mu_{i} \lesssim}$~20 mag~arcsec$^{-2}$ and effective radii of ${2\farcs1 \lesssim R_e \lesssim 9\farcs1}$, which corresponds to 165 to 740~pc at their intrinsic distances of, respectively, 16.3~Mpc \citep{Jordan2007} and 16.7~Mpc \citep{Blakeslee2009}. The properties of our sample galaxies are summarized in Table~\ref{tab:data}.

\begin{figure}
	\includegraphics[width=\columnwidth]{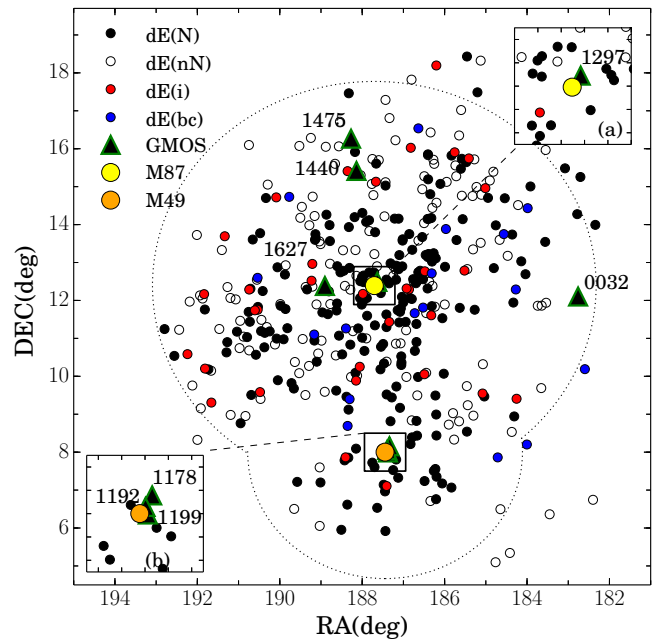}
    \caption{Location of our eight program galaxies within the Virgo Cluster. Our targets are indicated by the large black triangles with green contours and are labelled by their respective VCC numbers. Sub-panels (a) and (b) show magnified views of the one square degree regions centered on M\,87 (NGC\,4486) and M\,49 (NGC\,4472), respectively. A total of 413 ``dwarf'' ellipticals (dEs) from \citet{Lisker2007a} are plotted to show the global distribution of low-mass ETGs within the cluster. Colors and symbols correspond to their classification in sub-types (nucleated:~N, non-nucleated:~nN, disky:~i and blue core:~bc). The black dotted line indicates the virial radii, from \cite{Ferrarese2012a}, of the Virgo subclusters A (R$_{200}=1.55~Mpc$) and B (R$_{200}=0.96~Mpc$) centered on M\,87 and M\,49, respectively.}
	\label{fig:virgoloc}
\end{figure}

Although this was not a criterion for the selection, most of our objects are nucleated, or possibly nucleated, galaxies as shown in the ACSVCS survey \citep{Cote2004, Cote2006b}. They span a wide range in projected cluster-centric distance, from the center of the cluster close to M\,87 (NGC\,4486), out to Virgo's outskirts (see Fig.\ref{fig:virgoloc}). Three of them (VCC\,1178, VCC\,1192 \& VCC\,1199) have projected distances from M\,49 (NGC\,4472, the brightest galaxy of the Virgo cluster) of less than 50~\kpc\,, while another (VCC\,1627) is located close to the massive ETG M\,89 (NGC\,4552). A fifth galaxy, VCC\,1297 is a companion of the central massive ETG M\,87 (NGC\,4486). Redshift-independent distance estimates from surface brightness fluctuations \citep{Mei2007, Blakeslee2009} and globular cluster luminosity functions \citep{Jordan2007, Villegas2010}, show that the projected distances are representative of the intrinsic positions of the galaxies in the cluster, with the exception of VCC\,1627, which is located at the front-side of the cluster at a distance of 15.6~Mpc. VCC\,0032 has no redshift-independent distance, therefore we assume a mean Virgo distance of 16.5~Mpc. A more thorough discussion of the environment of our sample galaxies, and its possible influence on their structure, kinematic properties and stellar populations is presented in \S\ref{sec:discussion}.

\begin{figure*}
	\centering
    \includegraphics[width=\hsize]{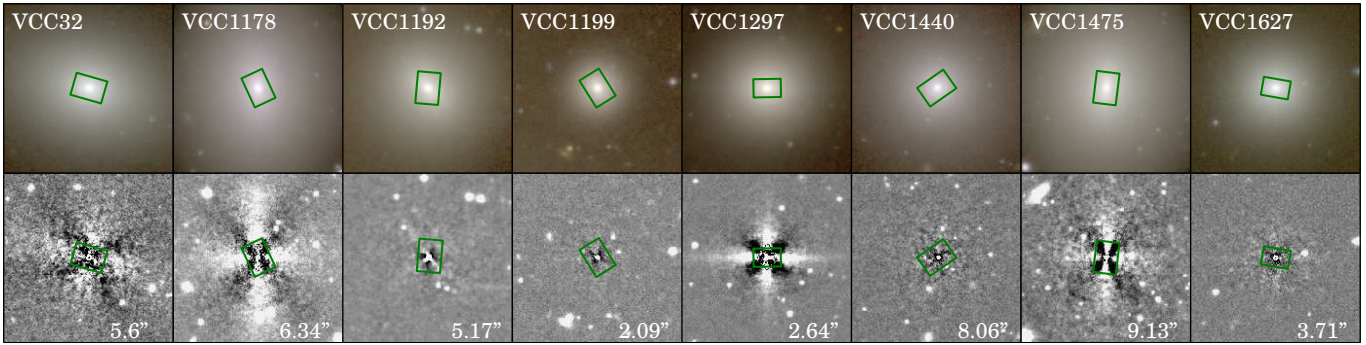}
	\caption{Composite \textit{giz}-colour images from the NGVS survey (top row) and NGVS \textit{i}-band residual images obtained by subtraction of a model computed with {\tt ELLIPSE}. The ellipse fit was performed by median-sampling of the image in elliptical annuli with semi-major axis incremented by 10\% at each iteration. During the procedure, all parameters (center, ellipticity, position angle) were allowed to vary. The models used to obtain the residual images shown in the lower panels do not include higher order moments and therefore only highlight deviations of the isophotes from pure ellipses. The green rectangle on each image indicates the \gmosifu\ FOV for the relevant mode (approximately $5\arcsec\times7\arcsec$ for 2-slit mode observations, and $5\arcsec\times6\arcsec$ for the mosaiced 1-slit mode observations). All images are oriented to show north-up and east-left. The effective radius of each galaxy is indicated at the bottom right of its residual image to give an idea of the relative coverage of our observations.}
	\label{fig:gmosFOV}
\end{figure*}

\subsection{Observations}
\label{sec:obs}
Our observations were carried out at the 8.1~m Gemini North Observatory, using the IFU of the Gemini Multi Object Spectrograph (GMOS, \citealt{Hook2004}), with a combination of one- and two-slit modes due to configuration availability. The observations were taken in service mode between March 2012 and July 2013 under the programs GN-2012A-Q-81, GN-2013A-Q-76 and GN-2013A-Q-108. Table~\ref{tab:dataobs} provides details on the instrumental setup and exposure times.

We used the \gmosifu\,\citep{Allington2002} primarily in its 2-slit mode, giving a $7\arcsec\times5\arcsec$ field-of-view (FOV) with 1000 hexagonal fibers of $0\arcsec.2$ projected diameter. 500 similar additional fibers, covering an area of $5\arcsec\times3\arcsec.5$ and pointing $\sim1$~arcmin away from the science field, are used to estimate the sky background level. Each slit covers half of the FOV and disperses over the corresponding half of the detectors. To avoid spectral overlap on the detectors between the spectra from the two slits, different filters can be set in front of the grating. We opted for the $B600$--$G5307$ grating combined with a {\textit{g}$^\prime$ filter}, giving a clean wavelength range of $3980$--$5480$\,{\AA}. This spectral range provides extensive coverage of strong stellar absorption lines (e.g., {\hbeta}, Fe and {\mgb} triplet) for measuring the kinematics as well as the population parameters of age and metallicity. The instrumental spectral resolution is 1688 at 4610\,\AA\ for the grating mentioned above, with an original sampling of 0.5\,\AA\ per pixel. We binned by a factor two in the spectral direction to get a higher signal-to-noise ratio (S/N) leading to an actual sampling of 1\,\AA\ per pixel and an instrumental spectral dispersion of about 75~\kms\ (or a FWHM of $2.7$\,\AA\ at 4610\,\AA).

Typical exposure times for our campaign were $4\times1800$ seconds and $6\times1800$ seconds: e.g.,~2 hours and 3 hours, respectively (see Table~\ref{tab:dataobs}), to reach a S/N of $20$ or better with spatial binning while keeping enough spatial resolution. One exception is VCC\,1178 for which an exposure time of $4\times900$ seconds was considered sufficient. Due to scheduling constraints on the \gmosifu\ configuration, two targets (VCC\,1297 and VCC\,1627) were observed in \gmosifu\ 1-slit mode. This mode covers only half of the FOV of the 2-slit mode, i.e.\,~$5\arcsec\times3\arcsec.5$, and as a consequence the time of integration had to be doubled to cover the original FOV. All other parameters are similar to the 2-slit mode except for an extended wavelength coverage, e.g.\,~$4200$--$7200$\,\AA\,. For consistency, however, we considered only the overlapping wavelength range between the 1-slit and 2-slit mode data in our analysis (i.e.\,, 4200--5480\,\AA).  

The GMOS focal plane consists of a mosaic of three $2{\rm k} \times 4{\rm k}$ e2vDD detectors with gaps equivalent to approximately 37 unbinned pixels each ($\sim20$\,\AA). To avoid losing spectral information, we used a spectral dithering technique by shifting the central wavelength of the grating by 50\,\AA\ for half of the exposures, using two central wavelength settings: 4800\,\AA\ and 4850\,\AA\,. Variations in the point spread function between exposures, however, complicates the optimal combination of the spectral dithers, so instead we simply excluded the spectral gaps when performing our analysis.

\begin{deluxetable}{ccccc}
 \tablewidth{0pt}
 \tablecaption{Instrumental Configuration
 \label{tab:dataobs}}
 \tablehead{
  \colhead{Galaxy} &
  \colhead{PA$_{\rm sky}$} &
  \colhead{PA$_{\rm inst}$} &
  \colhead{Run ID} &
  \colhead{T$_{\rm exp}$} \\
  \colhead{} &
  \colhead{(deg)} &
  \colhead{(deg)} &
  \colhead{} &
  \colhead{(sec)} \\
  \colhead{(1)} &
  \colhead{(2)} &
  \colhead{(3)} &
  \colhead{(4)} &
  \colhead{(5)} }
 \startdata
			VCC\,0032 & 76.0 & 164 & 1 & 4${\times}$1800\\
			VCC\,1178 & 5.2 & 295 & 2 & 4${\times}$900\\
			VCC\,1192 & 39.8 & 265 & 2 & 6${\times}$1800\\
			VCC\,1199 & 30.8 & 302 & 2 & 4${\times}$1800\\
			VCC\,1297 & 71.1 & 1 & 3 & 8${\times}$2100\\
			VCC\,1440 & 81.9 & 35 & 2 & 6${\times}$1800\\
			VCC\,1475 & -5.8 & 83 & 1 & 4${\times}$1800\\
			VCC\,1627 & -81.4 & 170 & 3 & 5${\times}$1800
\enddata
\tablecomments{(1) Galaxy name from the \citet{Binggeli1985} catalogue, (2)-(3) Position angle of the galaxy major-axis on the sky and GMOS instrument in the North-East frame, (4) Run: 1= GN-2012A-Q-81; 2 = GN-2013A-Q-76 (2-slit mode); 3 = GN-2013A-Q-76 (1-slit mode), (5) Total on-source exposure time in seconds.}
\end{deluxetable}

\subsection{Data reduction}
\label{sec:datareduc}

We used the official Gemini {\tt IRAF} package v1.11.1, released in March 2012, to reduce all our data: i.e.\,~perform bias subtraction, flat-fielding and illumination correction, wavelength calibration, cosmic ray-rejection and sky-subtraction. The final cubes (formed from a combination of individual exposures) were created using the {\tt XSAURON} software \citep{Bacon2001SAURONI}, developed by the \textit{Centre de Recherche Astrophysique de Lyon} (CRAL). We developed a dedicated {\tt PYRAF} script to automate the end-to-end data reduction, ensuring a homogeneous treatment of the reduction procedure across all targets.\

Raw frames (e.g.\ science, flat lamp and twilight exposures) were overscan and bias subtracted using the task \textit{gfprepare}. The arcs were only overscan subtracted as they were taken in fast-read mode. The dark current of GMOS e2vDD CCDs is typically less than 1 electron per hour and was thus ignored.\

The flat lamps within the Gemini calibration unit, GCAL, were used to derive the flat-field response map for each associated science exposure. We used the task \textit{gfreponse} to derive the maps from the flats and we used one twilight per observation run (see Table~\ref{tab:dataobs}) to correct for illumination. The science spectra were extracted using the aperture mask derived from their associated flat lamp (taken immediately before or after the exposure) and corrected for flat-fielding and illumination using the task \textit{gfextract}. As explained in the \S~\ref{sec:obs}, in 2-slit mode each slit disperses over one half of the detector mosaic and a filter ($g^\prime$ in this case) was used to avoid overlap between the spectra coming from each slit. However, the blue end of the spectra coming from the left side of the FOV can still overlap with the red end of the spectra coming from the right side of the field. Thus, the spectra were shortened by 10\% ($\sim12$\,\AA) in the overlapping area to eliminate any contamination from one slit on the other.\

We checked the quality of the flat-fielding by applying our procedure on twilight exposures, which should provide a homogeneously illuminated field. We found a residual uncertainty of about 1\% between spectra from the same slit, but a systematic offset of $10-20$~\% between the two slits. This issue is known to arise from errors in the normalization from the Gemini {\tt IRAF} flat-fielding recipe itself and not from the data. We chose not to correct for this offset since our analysis does not require accurate absolute flux calibration, and the spectral shape is not significantly affected. However, the isophotes derived from the integrated flux in the data cubes show some residual asymmetry as a result of this effect (see Fig.~\ref{fig:maps}), and so should be viewed with caution.

We derived the wavelength solution based on the GCAL arc lamp exposures using the task \textit{gswavelength} with a 4th-order Chebyshev polynomial. Due to charge transfer efficiency issues that can systematically affect the relative centroid of strong and weak lines, as well as introduce effects related to the read-out direction of the detector, we fitted only the 20 brightest lines available in the CuAr arc lamps. We carefully checked that those lines sample the whole wavelength range appropriately. Unfortunately, no bright lines are available redwards of 5187\,\AA\ when using a central wavelength of 4800~\AA\, which reduced the accuracy of our wavelength solution beyond that wavelength for that configuration. We therefore excluded the wavelength range $5180$--$5480$\,\AA\ for the kinematic analysis. Unfortunately, the \mgb\/ triplet lines were consequently excluded from the fit for the majority of our galaxy sample, but the S/N of our spectra is sufficiently high to still allow us to reach the low velocity dispersion levels required (see \S\ref{sec:binning} and Fig.~\ref{fig:mcsim} for details). We obtained typical formal RMS errors in the wavelength calibration of $0.1$\,\AA\ and $0.08$\,\AA\ for the 4800~\AA\, and 4850~\AA\, central wavelengths settings, respectively.\

Finally, we ran the task \textit{gscrrej} on each exposure to reject cosmic rays and cleaned the remaining bright cosmic impacts via another cosmic ray rejection algorithm implemented within {\tt XSAURON}. The sky contribution in the science spectra was subtracted by running the task \textit{gfskysub} using the median of all dedicated sky fibers.\

\section{Analysis}
\label{sec:analysis}

\subsection{Merging \& Binning}
\label{sec:binning}
\begin{figure}
    \includegraphics[width=\columnwidth]{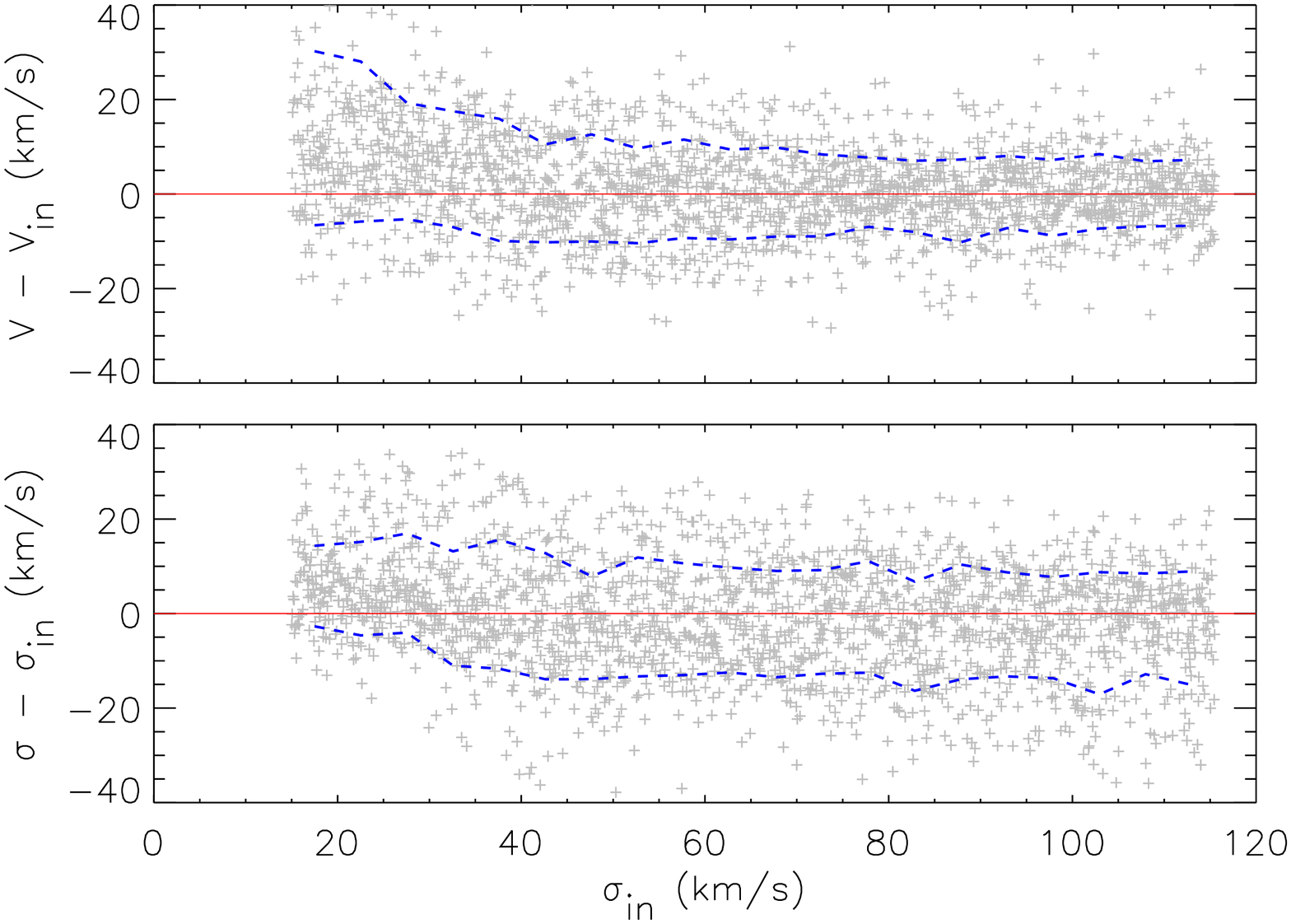}
	\caption{Monte-Carlo simulation of \gmosifu\ observations, made using a model spectrum of a 12.59\,Gyr, solar metallicity population from \citet{Vazdekis2012aMIUSCAT}. The spectral resolution, sampling and wavelength range have been matched to our observations. The S/N was set to 25. Gray points indicate the 2000 individual simulation results. Blue lines show where 68\% of the points lie, as calculated within bins of at least 100 simulations, giving an estimate of the uncertainties. Errors on mean velocity, $V$, and velocity dispersion, $\sigma$, are less than 11~\kms\/ for velocity dispersions above 35~\kms. For dispersions below this value, the recovered parameters become biased to some degree, indicated by the discrepancy in recovered $V$ values.}
	\label{fig:mcsim}
\end{figure}

We used the software {\tt XSAURON} to reconstruct the final cubes starting from the individual reduced exposures described in \S\ref{sec:datareduc}. All individual exposures were first truncated to a common wavelength range ($4200$--$5400$\,\AA) and spatially aligned using isophote contours. The exposures were then merged by interpolating each one onto a common $0.2\arcsec\times0.2\arcsec$ grid, and then taking the optimally weighted sum of the exposures at each pixel.

To increase the S/N of our data, we used the adaptive spatial binning software developed by \cite{Cappellari2003Bin}, based on a Voronoi tessellation. The formal error spectra are not fully propagated through the reduction process by the Gemini {\tt IRAF} package. Therefore, as a first approximation, we assumed that the noise in a given spaxel was directly proportional to the square root of its integrated flux (i.e. Poissonian). We performed a spectral fit of the central unbinned spectrum (see \S\ref{sec:kinextraction} for a description of the spectral fitting method used) to get a good estimate of the noise from the fit residuals, which we used to renormalize our Poissonian noise estimate before applying the Voronoi binning. We tested that this renormalization did not depend on which region of the galaxy was used to compute the fit, and found no dependence.

ETGs in the luminosity range of our program galaxies have typical velocity dispersions of $\sim$30~\kms, which is well below our instrument dispersion of 75~\kms. We used Monte-Carlo simulations with model spectra adapted to match the spectral properties (wavelength range, resolution, pixel size) of the \gmosifu\, observations. Fig.~\ref{fig:mcsim} shows that, at velocity dispersions of 30--40~\kms,  binning our data to a S/N threshold of 25 was sufficient to recover low velocity dispersions with errors of $\sim 11$~\kms\/ for both the mean velocity $V$ and velocity dispersion $\sigma$. This was considered an acceptable compromise between the kinematic errors and degree of spatial binning. During the binning process, we rejected all spectra with a S/N lower than 3 as to avoid creating very large poor quality bins. The adaptive spatial binning on average delivers the target S/N in each spectrum within a standard deviation of 8\%. From our simulations, this small spread in S/N corresponds to a variation of only 1--2~\kms\,~in the errors of velocity and velocity dispersion, and similarly small relative errors on the line strengths.

\subsection{Kinematics extraction}
\label{sec:kinextraction}
To measure the stellar kinematic properties of our objects (mean radial velocity, $V$, and velocity dispersion, $\sigma$), we fitted the data using the penalized pixel-fitting method (pPXF) from \cite{Cappellari2004Ppxf} as implemented in {\tt XSAURON}. We used an empirical library of 89 stars taken from the MILES library \citep{SanchezBlazquez2006MILES, Falcon2011} that covers the spectral range of $3525$--$7500$\,\AA\ at a spectral resolution of 2.3\,\AA\ FWHM ($\sigma \sim$60~\kms\ at 5000\,\AA), similar to (but lower than) the \gmosifu\ spectral resolution. We selected these stars in such a way to cover uniformly the parameter space (T$_{eff}$, log(g), [Fe/H]) expected for dwarf galaxies. For each individual target, we first fitted the spaxel having the highest S/N to obtain the combination of stars that best reproduces the galaxy spectrum. We then used this single best-fit template for all spaxels to minimize the impact of template variations within the FOV. We fitted the spectra over the range $4200$--$5180$\,\AA\,, excluding the redder wavelengths for the reasons mentioned in \S\ref{sec:datareduc}.

We performed the same kinematic extraction on the twilights obtained during the different runs (observed with the same setup and reduced in the same way as the science exposures) to check the quality of our wavelength calibration. Ideally, the twilight illuminates the full \gmosifu\ FOV and one can expect to obtain the same radial velocity and velocity dispersion on each of the 1000 science fibers. Any variations from one fiber to another will represent the uncertainties of our measurements due to, e.g.,\,~systematic instrumental signatures, reduction processes, etc. In 2-slit mode, we measured a small radial velocity systematic offset of $\sim$5~\kms\, between the two slits of the \gmosifu\ instrument (i.e.\ the two sides of the field), and variations of a few \kms\, within a single slit (for both 1-slit and 2-slit modes). Both measurements are within the wavelength calibration error (rms $\sim$0.1\,\AA\,, i.e.\,,~$\Delta\,V\sim$7~\kms) and kinematic errors due to noise as estimated in Fig.~\ref{fig:mcsim}. 

We measured on the final binned maps (see \S~\ref{sec:kin} and Fig.~\ref{fig:maps}) the standard deviation of the radial velocity V and velocity dispersion $\sigma$~over a 10$\times$10 spaxels area where the binning was strong (i.e.\,,~at the edge of the \gmosifu\,~FOV). We performed these measurements on objects having relatively flat V and/or $\sigma$ profiles (e.g.\,,~VCC\,0032, VCC\,1440, VCC\,1627) so these quantities represent the global true uncertainties on the kinematics. The typical standard deviations found are of 10~\kms\,~for V and 15~\kms\,~for $\sigma$, which are similar to the values estimated from our simulations.

\subsection{Age and metallicity measurements}
\label{sec:ageZmeasurement}
To study the stellar populations in our galaxies, we apply spectral fitting to derive values of age and metallicity using the MIUSCAT models \citep{Vazdekis2012aMIUSCAT} that provide single age, single metallicity stellar population models based on MILES and CaT \citep{Cenarro2001} libraries. We used a subset of 203 models covering an age range of $0.5$--$12.5$~Gyr, a metallicity range of $-2.32$ to +$0.22$~dex, and computed using a unimodal IMF with a slope of 1.3, equivalent to a \cite{Salpeter1955} IMF. The model spectra were retrieved from the MIUSCAT website, sampled at 2.5\,\AA\ FWHM resolution (similar to \gmosifu\ data) and covering the wavelength range $3464$--$5550$\,\AA\,. As with the kinematics extraction, we used pPXF software to fit the galaxy spectra over the wavelength range of $4200$--$5400$\,\AA\,, excluding \hbeta\, and [OIII] lines to avoid being contaminated by any possible emission from ionized gas.\

\subsubsection{Regularized spectral fits}
\label{sec:regulmethd}
As our primary approach, we used regularization of our spectral fits such that the best distribution of templates found to fit the data is also the smoothest in the parameter space (in this case, age and metallicity). In other words, we impose the condition that the star formation history is the smoothest possible. The age and metallicity values are then derived by taking the mean age and metallicity of the selected templates, weighted by their mass contribution to the combined template. This is equivalent to the spectral fitting approach used by \cite{McDermid2015} for the \atlas\, survey.\

We performed this analysis on each spaxel of our galaxy cubes, spatially binned to a S/N of 35. This S/N value was chosen to preserve spatial information on each target while still providing robust measurements and acceptable errors. Two targets, VCC\,1199 and VCC\,1627, were kept with a S/N of 25 as the data are of lower quality and we opted to retain the smaller bin size. For each individual target, we used the spaxel with the highest S/N to derive the degree of regularization to use for all the spaxels. In this way, we keep the degree of regularization constant over the full FOV. The maps of the mass-weighted age and metallicity are shown in Fig.~\ref{fig:maps} and discussed in \S\ref{sec:results}.\

To estimate the errors made on the mass-weighted age and metalliticy, we performed the same exercise on the stellar population binned maps as with the kinematic maps and found typical standard deviations of 0.4~Gyr in age and 0.05~dex in metallicity. However, it is important to notice that these measurements are mostly formal uncertainties which mostly depend on systematics and the used spectral library. The exact impact of the library used for the measurements is beyond the scope of this paper.

\subsubsection{Best single SSP fit}
\label{sec:sspmethd}
In order to compare the stellar population properties of our sample galaxies with others in the literature \citep{Koleva2011, Rys2014, Toloba2014bStellPop, Toloba2014cLR}, we also carried out a single stellar population (SSP) analysis similar to the spectroscopic indices method \citep{Michielsen2007, Koleva2011aSSPindices}. This method, which is well studied and described in \cite{Koleva2008}, consists of selecting the single SSP model that, in terms of $\chi^2$, best fits the data, taking the age and metallicity of this model as the measured value. To reduce the limitations of discreteness in the grid of model age and metallicity values, we interpolated the $\chi^{2}$ surface within the parameter space log(age)-[Z/H]. We used this interpolation to determine the position of the $\chi^{2}$ minimum and the $\Delta \chi^{2}$ error surface.\

We performed this analysis for each of our targets on the single spectrum obtained by spatially stacking spectra within an ellipse equivalent in area to a circular aperture of $R_{e}/2$, except for two targets, VCC\,1440 and VCC\,1475, that we only cover up to $R_{e}/3$. Effective-ellipses have been derived using moments of the surface brightness of our galaxies in the NGVS \textit{i}-band images. The corresponding ellipticities are given in Table~\ref{tab:data}.

\begin{figure}
    \includegraphics[width=\columnwidth]{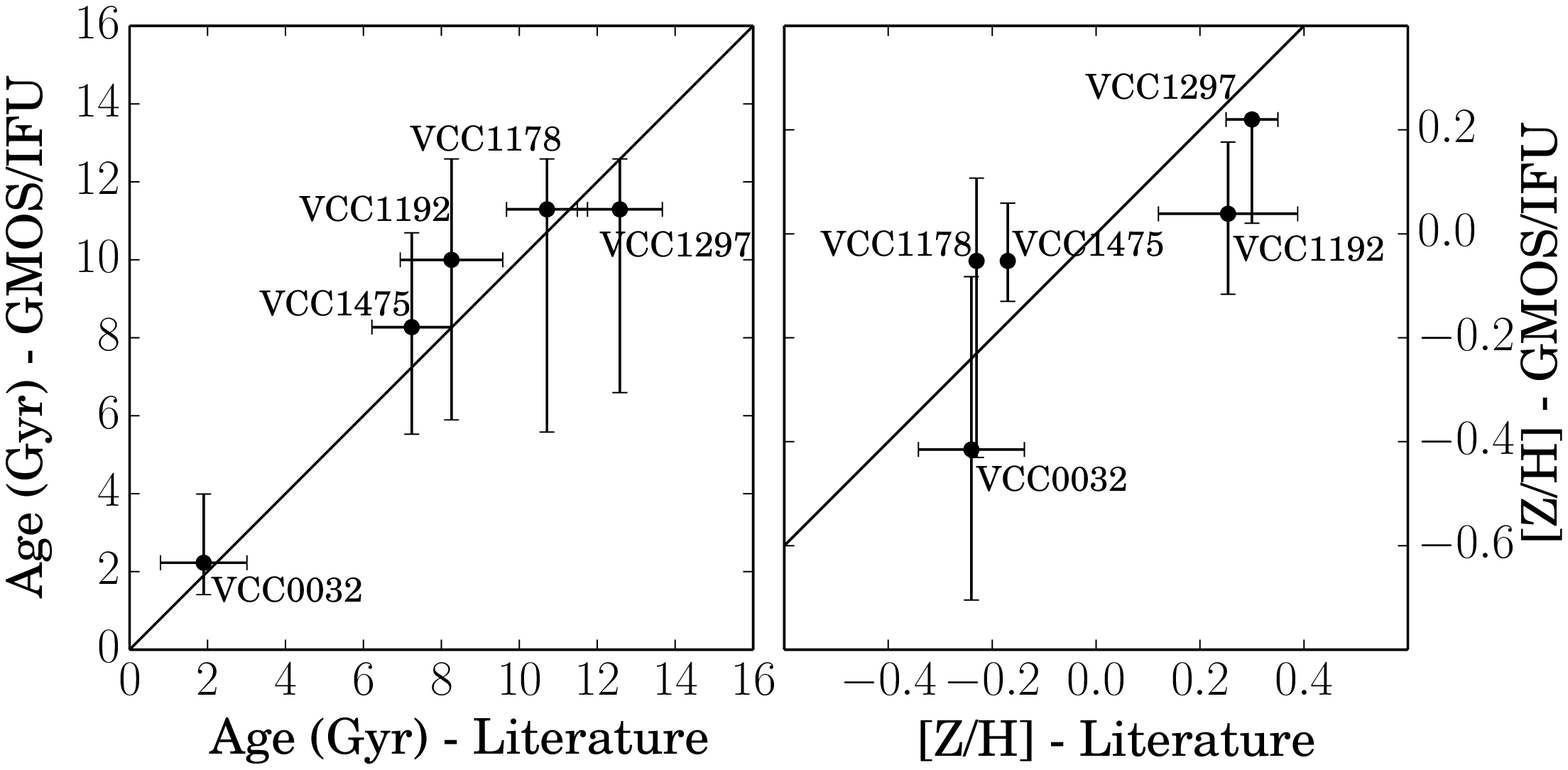}
	\caption{Comparison of SSP age and metallicity values from our \gmosifu\, survey and \citealp{Sanchez-Blazquez2006b} (VCC\,0032, VCC\,1192), \citealp{Koleva2011} (VCC\,1178, VCC\,1475) and \citealp{Spolaor2010b} (VCC\,1297). For the \gmosifu\ values, the error bars were derived from the 1$\sigma$ contour of the marginalized $\chi^{2}$ surface (i.e.\,~$\Delta\chi^{2}=1$) in the log(age)-[Z/H] parameter space. The upper age error of VCC\,1178, VCC\,1192 and VCC\,1297 are set by the boundary of the models used, i.e.\,,~12.5~Gyr.}
	\label{fig:complit_ageZ}
\end{figure}

We compared our results with published values for VCC\,0032 and VCC\,1192~\citep{Sanchez-Blazquez2006b}, VCC\,1178 and VCC\,1475~\citep{Koleva2011} and VCC\,1297~\citep{Spolaor2010b} by simulating their long-slit data and performing the analysis on the same apertures. Our findings are in good agreement with these previous studies, as illustrated in Fig.~\ref{fig:complit_ageZ}. The errors on our measurements are taken to be where the marginalized $\chi^{2}$ surface of the fit shows a difference of 1 with respect to the minimum value, corresponding to the 68~\% confidence limits.\

\section{Results}
\label{sec:results}
\begin{figure*}
	\centering
	\includegraphics[scale=0.88]{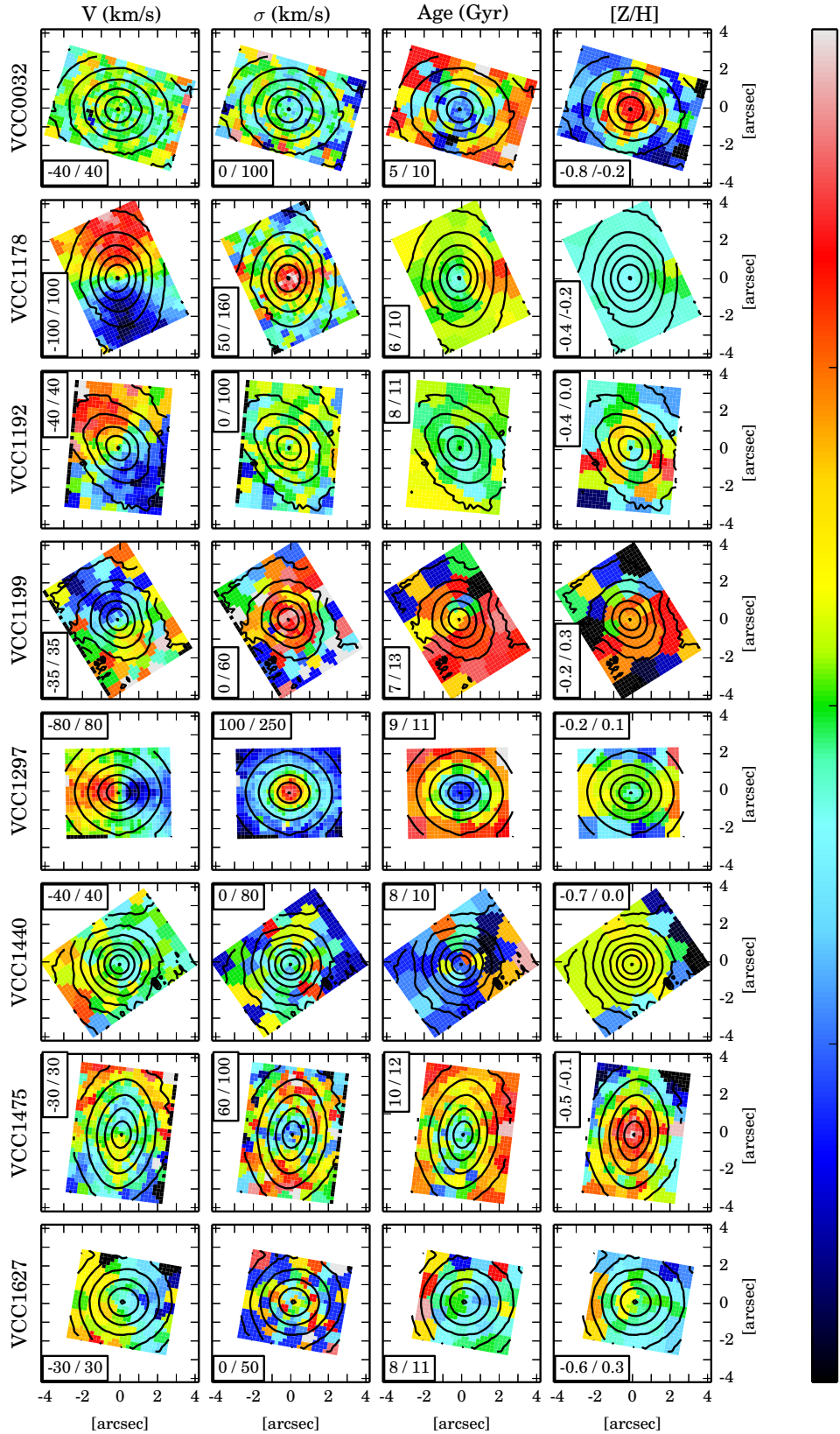}
	\caption{Maps of mean radial velocity, $V$, velocity dispersion, $\sigma$, and stellar population parameters (age and metallicity [Z/H]) for our eight low-mass ETGs in the Virgo cluster. The box in each sub-panel indicates the values corresponding to the extreme colors of the colorbar on the right. The dark contours in each panel represent the iso-photometric contours of the specific target computed from the integrated reconstructed cube.}
	\label{fig:maps}
\end{figure*}

Several IFU-based studies of low-mass ETGs, spanning a broad range of properties, have been published in recent years \citep{Prugniel2005, Chilingarian2009a, Rys2013}. As discussed in \S\ref{sec:sample}, our study focuses on the most compact low-mass ETGs, thereby probing a largely unexplored regime in size ($165 \lesssim R_{e} \lesssim 740$~pc) and surface brightness $\mu_{i} \lesssim 20$~mag~arcsec$^{-2}$. From an observational perspective, such objects are challenging targets due to their small sizes; as a result, only a few have had their basic kinematic and stellar population properties characterized. The spatial resolution and sensitivity of the \gmosifu\ have allowed us to obtain resolved two-dimensional spectroscopy of the central parts of eight of these compact objects in the Virgo cluster. The kinematics, age and metallicity maps for this sample are presented in Fig.~\ref{fig:maps}. We also provide some further details pertaining to two noteworthy objects (VCC\,1297 and VCC\,1475).\

\subsection{Kinematics}
\label{sec:kin}

Six of the eight galaxies in our sample show significant stellar rotation. The exceptions, VCC\,0032 and VCC\,1440, have mean stellar rotation velocities consistent with zero within the entire \gmosifu\ FOV: i.e.\,,~up to about half the effective radius for VCC\,0032, and a third of the effective radius for VCC\,1440. Interestingly, these two galaxies happen to reside in the outskirts of the cluster (see Fig.~\ref{fig:virgoloc}). Typical mean radial velocity amplitudes for the other six objects are between 15 and 30~\kms, which is similar to other low-mass, but more diffuse, ETGs  \citep{Rys2013, Toloba2011}. Two galaxies, namely VCC\,1178 and VCC\,1297, show significantly larger stellar velocities (90 and 68~\kms, respectively). Our \gmosifu\ data also confirm and highlight the presence of a counter-rotating core in VCC\,1475. This feature was first noted by \cite{Koleva2011} using long-slit data and later re-analyzed by \cite{Toloba2014a} (see \S\ref{sec:VCC1475} for more details).

The galaxies in our sample have typical stellar velocity dispersions of 20--60~\kms, again similar to what has been observed for other low-mass ETGs. Half of the sample has fairly flat stellar dispersion profiles (VCC\,0032, VCC\,1192, VCC\,1440 and VCC\,1627). However, the remaining four galaxies show significant central dispersion peaks, rising to values of 60~\kms\ (VCC\,1199), 70~\kms\ (VCC\,1475), 155~\kms\ (VCC\,1178) and 235~\kms\ (VCC\,1297). The last three objects also show significant organized rotation, as well as disky isophotes in the photometry of the same central regions (visible in Fig.~\ref{fig:gmosFOV}.)\

Fig.~\ref{fig:gmosFOV} shows the composite \textit{giz}-images from the NGVS survey as well as the \textit{i}-band residual images of our target galaxies, obtained by subtracting from the original image the best ellipse fit at each radius (from the center and incremented by 10\% at each iteration). The center, ellipticity and position angle of the fit were allowed to vary but the models used to obtain the residual images do not include higher order moments and therefore only highlight deviations of the isophotes from pure ellipses. Three galaxies (VCC\,1199, VCC\,1440 and VCC\,1627) show little structure in their residual images, but typical ``X-shape" residual of disky isophotes are apparent in the other five (VCC\,0032, VCC\,1178, VCC\,1192, VCC\,1297 and VCC\,1475). This is confirmed by the cosine fourth order moment in the Fourier expansion of the intensity \citep{Jedrzejewski1987} which is positive out to several tens of arcsec for VCC\,0032, VCC\,1778, VCC\,1297 and VCC\,1475, and out to a few arcsec for VCC\,1192. This is in agreement with the potential high degree of substructure present in low-mass ETGs (e.g., \citealt{Lisker2007a, Turner2012}). Note that the lack of structure in the residual images can also be due to projection effects, at least for VCC\,1440, which is a round object ($\epsilon$~$\sim$~0.013) that does not show much rotation, and therefore could be an intrinsically flattened object seen close to face-on.\

A significant fraction of the sample of dwarfs observed by \cite{Chilingarian2009a} exhibited central drops in velocity dispersion \citep[see also][]{Rys2013}. In our sample, only VCC\,1475 shows such a decreasing velocity dispersion profile with decreasing radius. This feature is certainly linked with the presence of the two counter-rotating components, which induce two peaks in the velocity dispersion map along the major-axis, rather than a decrease in the center (see \S\ref{sec:VCC1475}).\

Although our small sample is not a complete census of the compact low-mass ETGs in the Virgo cluster, {\it it is quite striking that we observe such a variety of kinematic signatures}. Indeed, the apparent diversity is {\it remarkably similar to what is observed in more massive ETGs: from non- to highly-rotating systems, and the existence of large-scale, kinematically decoupled components.} 

Further details regarding the dynamical support of the galaxies in our sample are provided in \S\ref{sec:discussion} via the specific stellar angular momentum parameter, $\lambda_{R}$, defined in \cite{Emsellem2007LR}.\

\subsection{Stellar populations}
\label{sec:stellarpop}
Mapping the stellar population parameters of age and metallicity, along with kinematic substructure, can provide interesting constraints on the formation and assembly history of galaxies. Here we present our results on the mass-weighted age and metallicity distributions of the central regions of our objects, obtained as described in \S\ref{sec:regulmethd}. Our SSP-equivalent age and metallicity estimates are presented in Table~\ref{tab:dataresults} and discussed in more detail in \S\ref{discussion_stellpop}.\

On average, our objects are comprised of old to very old stars, with mass-weighted mean ages of 6--11.5~Gyr. The age variation within one galaxy spans a relatively large range from 0--3.5~Gyr between the center of the galaxy and the maximum radius covered by \gmosifu\ FOV, i.e.\,~about half of the effective radius. Indeed, four galaxies do not show any age gradient (VCC\,1192, VCC\,1199, VCC\,1440 and VCC\,1627) whereas the four others show negative variation of 0.5~Gyr (VCC\,1178), 1~Gyr (VCC\,1475), 1.5~Gyr (VCC\,1297) and 3.5~Gyr (VCC\,0032) going toward the center (i.e., younger in the core). Younger central stellar populations in low-mass ETGs were previously noted by others \citep{Michielsen2008, Chilingarian2009a, Paudel2011, Rys2013} and are interestingly present in both non-rotating (e.g.\,~VCC\,0032) and rotating (e.g.\,~VCC\,1297) objects.\
In terms of metallicity, the results of the regularized full spectral fitting show a wide range in behavior among our objects, from fairly flat profiles (e.g.\,VCC\,1178, VCC\,1192, VCC\,1199, and VCC\,1440) to positive gradients with decreasing radius (e.g.\,VCC\,0032, VCC\,1297, VCC\,1475 and VCC\,1627). All our objects have solar to sub-solar stellar populations with metallicity, [Z/H], ranging from $-0.8$ to $+0.2$~dex, with a typical variation of, at most, 0.2~dex between the center of the galaxy and the edge of the \gmosifu\, FOV. In most cases, the central regions, which are composed of more metal-rich stars, correspond to younger stellar populations (see Fig.~\ref{fig:maps}), in agreement with \cite{Michielsen2008} and studies of more extended dwarf galaxies in the Virgo cluster~\footnote{These values are also consistent with the well studied compact elliptical M\,32 (NGC\,221) which shows a solar mean metallicity and a mass-weighted age of 6.8~Gyr at $\sim$~2.7 effective radius \citep{Monachesi2012} for an effective radius of $R_{e}\sim$~100~pc \citep{Graham2002} and an estimated dynamical mass of 8$\times$~$10^{9}$~\Msun\,~\citep{Cappellari2006}. Its kinematics is also similar to our \gmosifu\, objects, having a large-scale radial velocity amplitude of about 40~\kms\,~\citep{Howley2013} and a velocity dispersion $\sigma_{Re}\sim$~60~\kms\,~\citep{Cappellari2006}.}. However, VCC\,1297 has an even more interesting stellar population structure that is discussed below.

\subsection{Specific cases}
\label{sec:SpecCase}

\subsubsection{VCC\,1297: a central black-hole}
\label{sec:VCC1297}
\textit{VCC\,1297 (NGC\,4486b)} is a compact, low-mass ETG that lies very close to the center of the Virgo cluster, just 35~\kpc\, from M\,87 in projected distance. This M\,87 companion galaxy has been well studied and is thought to have a central black hole of $6\times10^{8}$~\Msun\,\citep{Kormendy1997}. It has been reported to show a double nucleus with a separation of 0\farcs15 \citep{Lauer1996}, but this feature was not evident in imaging from the ACSVCS \citep{Ferrarese2006,Cote2006b}. Our observations cover the galaxy up to its effective radius of 2\farcs3. Despite its fairly round isophotes ($\epsilon=0.088$), a thin, fast-rotating disk-like structure is observed in the central 3\arcsec$\times$1\arcsec\, region that corresponds to the location of the disky isophotes revealed by the NGVS residuals image (see Fig.~\ref{fig:gmosFOV} and also \citealt{Ferrarese2006}). Its stellar radial velocity peaks at 68~\kms\ at $\sim$1\arcsec\, along the major axis, in agreement with \cite{Kormendy1997} to within the uncertainties. The \gmosifu\ data also reveal that the galaxy rotates much more slowly ($\sim$20~\kms) at $\sim1$\arcsec\,~away from the major-axis, similar to its rotation at larger radii ($r\sim4$\arcsec) along its major-axis. The velocity dispersion is about 150~\kms\ within one effective radius but peaks, after a steep inward rise, at 235~\kms\ within the central 1\arcsec$\times$1\arcsec\,. This strong peak in velocity dispersion leads to a S\'{e}rsic-dependent dynamical mass of $1.1\times10^{10}$~\Msun~(see \S~\ref{sec:galdensity} for details), implying that the black hole contributes more than 5\% of the dynamical mass. This puts it significantly above the fundamental relation of black hole -- host galaxy masses \citep{Ferrarese2006}.

\begin{figure*}
	\includegraphics[width=\hsize]{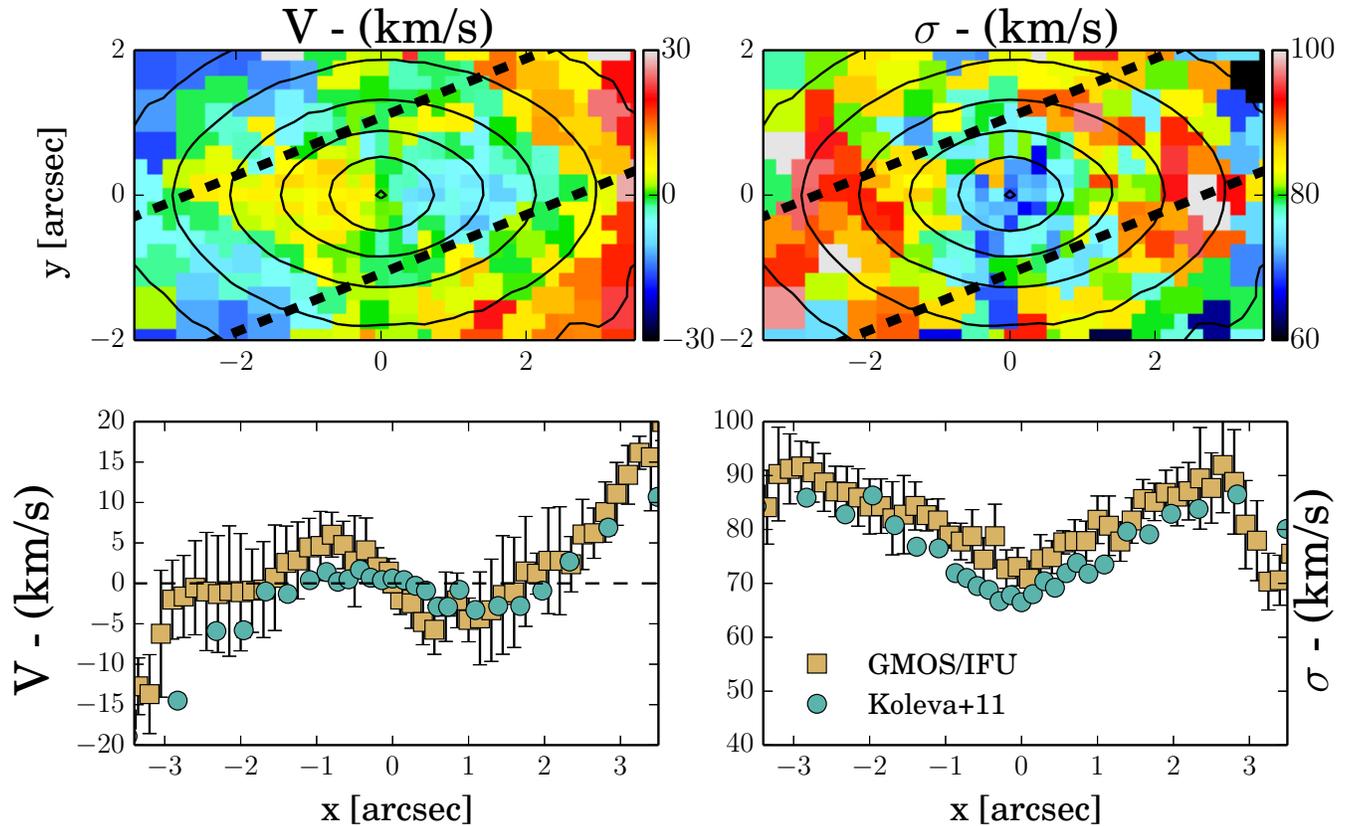}
    \caption{Radial stellar velocity, V, and velocity dispersion, $\sigma$, maps for VCC\,1475 from \gmosifu\ data (top row). The x and y axes are both in arcseconds. The black dotted lines indicate the region of the galaxy observed with a long-slit by \citet{Koleva2011}. The bottom panels show comparisons of the \citet{Koleva2011} $V$ and $\sigma$ profiles with comparable long-slit data simulated from our \gmosifu\ observations. The error bars are computed using the light weighted standard deviation of the points considered within the simulated slit. \citet{Koleva2011} error bars are typically of 0.7~\kms\, and are smaller than the marker points.}
	\label{fig:VCC1475}
\end{figure*}

Very interestingly, the increase of velocity dispersion at the galaxy center coincides with a drop in age. The stars in the central regions are found to be 1.2~Gyr younger  than the main body of the galaxy (i.e., 9.3 versus 10.5~Gyr). At the same time, the metallicity map shows a 0.1 dex enhancement along the fast-rotating disk-like structure compared to the main body (0.0 versus --0.1~dex). A gradient within the disk is also visible, with the nucleus being less metal-rich than its outer part. These observations suggest that an episode of star formation happened at the center of this galaxy after its formation, possibly triggered by an external gas accretion that would also explain the metallicity gradient within the disk (i.e.\,~more metal poor in the center).  

\subsubsection{VCC\,1475: a $2-\sigma$ galaxy}
\label{sec:VCC1475}

\textit{VCC\,1475 (NGC\,4515)} is a low-mass ETG containing a kinematically decoupled core (KDC) that was first identified by \cite{Koleva2011} using long-slit data and later re-analyzed by \cite{Toloba2014a}. This galaxy is located in the outskirts of the cluster, about 1.1~\mpc\,~in projected distance from M\,87 (i.e.\,, 0.7 the subcluster A virial radius). It has a mildly flattened shape ($\epsilon\simeq0.2$) within its effective radius of 9\farcs3 but becomes increasingly flattened ($\epsilon\simeq0.42$) within its central 2\arcsec$\times$2\arcsec\,, resulting in an ellipticity of 0.35 within the \gmosifu\ FOV. The isophotes in that region are disky, as emphasised in the residual image presented in Fig.~\ref{fig:gmosFOV}. Our two dimensional data cover the full KDC and reveal a true counter-rotating disk (aligned with the major-axis) that has a minimal size of $\sim 5$\arcsec\, (i.e.\,~$\sim 400$~pc) in diameter. The two stellar counter-rotating components have a mean velocity amplitudes of 8~\kms\ (inner component) and 16~\kms\ (outer component, within the \gmosifu\,~FOV). The stellar velocity dispersion map shows two off-centered peaks along the major-axis, making VCC\,1475 a bonafide $2-\sigma$ galaxy \citep{Krajnovic2011}. The amplitudes of the peaks are of order 100~\kms\ and correspond to the apparent transition regions between the two counter-rotating disks. The velocity dispersion drops to 70~\kms\ in the nucleus. We thus confirm the discovery of a KDC in VCC\,1475 by \cite{Koleva2011}, but clearly reveal its counter-rotating and $2-\sigma$ nature with our \gmosifu\ data. Figure \ref{fig:VCC1475} presents a magnified view of the kinematic maps for VCC\,1475 as well as a comparison of the rotation curve of \cite{Koleva2011} with that extracted from our \gmosifu\ data.\

Like VCC\,1297, this $2-\sigma$ galaxy has a stellar population that is 1~Gyr younger in its core region than in the external region covered by our observations (9.5 versus 10.5~Gyr). This outward increase in age is accompanied by a decline in metallicity going from the central to outer regions (--0.1 versus --0.5~dex, respectively). These differences in stellar populations seem to spatially match the two kinematically distinct components and are very similar to the behavior seen in VCC\,0032, both in terms of age and metallicity gradients. The kinematically distinct inner component in VCC\,1475, coupled to its younger stellar populations, strongly suggests that a dissipative accretion event has occurred in this galaxy, followed by an episode of star formation.

\section{Discussion}
\label{sec:discussion}

In \S~\ref{sec:analysis} and \S~\ref{sec:results}, we have analyzed two-dimensional spectroscopic data for eight compact dwarf ellipticals in the Virgo cluster obtained with the \gmosifu\ instrument. In this section, we discuss our findings in the context of possible ETG evolution scenarios.

\subsection{Mass, size and local galaxy density}
\label{sec:galdensity}

A key goal of our study is to understand whether the compactness of our program objects is related in any way to a specific physical evolutionary process. Many previous investigations have emphasized the importance of harassment \citep{Moore1998, Boselli2008a}, strangulation \citep{Smith2012}, ram pressure and tidal stripping \citep{Gnedin2003a, Mayer2006, Lisker2013, Toloba2014cLR} and, more generally, interactions with the underlying galaxy population or inter-galactic medium, in the evolution of low-mass galaxies \citep{Boselli2014}. In that spirit, we first turn our attention to the relationship between the basic properties of our targets and the galactic density in the regions in which they reside. 

\begin{figure*}
	\includegraphics[width=\hsize]{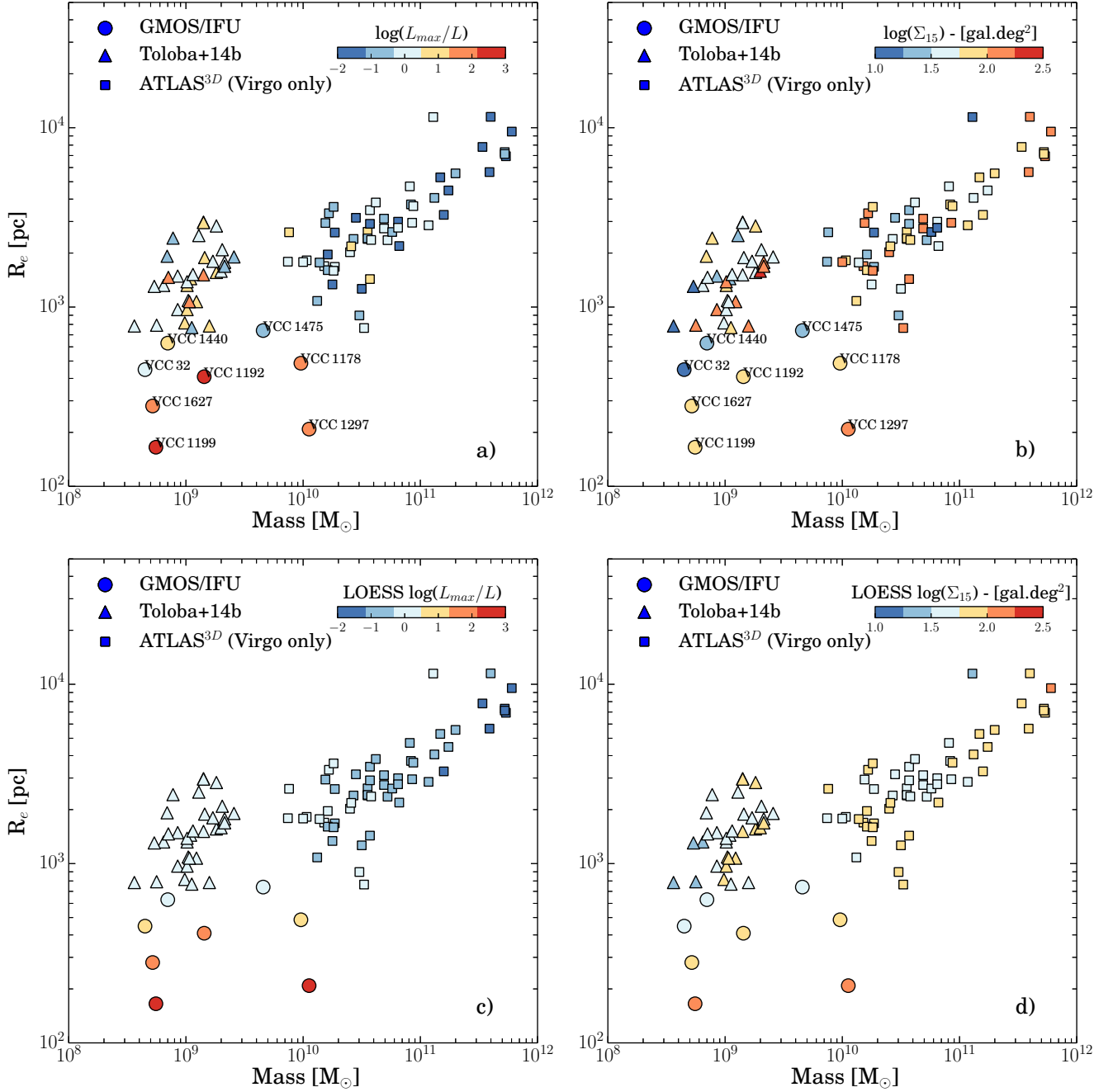}
    \caption{Location of Virgo galaxies from our \gmosifu\, sample, those of \cite{Toloba2014bStellPop} observed with long-slit, and \cite{Cappellari2013XX} from the \atlas\, survey, in the mass-size ($M-R_e$) plane, with color-coded symbols representing a proxy of the local galaxy density: the \textit{a} and \textit{c} (left) panels use the luminosity of the brightest galaxy among the 15 closest neighbors of the target galaxy, normalized by its luminosity, $L_{max}/L$ ; the \textit{b} and \textit{d} (right) panels use the $\Sigma_{15}$ parameter defined as the number of galaxies per square degree within a region that includes the 15 closest neighbors to the target galaxy. The bottom panels show a LOESS version of the original dataset presented in the top panels using a regularization factor of 0.3. The gap between the low- and high-mass ETGs is due to sample selection effects.}
	\label{fig:sizemassDensity}
\end{figure*}

In Fig.~\ref{fig:sizemassDensity}, we use the unique coverage and depth of the NGVS dataset to examine this relationship for Virgo galaxies using a mass-size diagram (i.e., effective radius, $R_e$, versus dynamical mass, $M$). We compute masses for our objects using the virial theorem,
\begin{equation}	
	M\,=\,0.7\,\beta(n)\,\sigma{_{R}}{^{2}}\,R\,/\,G
\end{equation}
where $\beta(n)$ depends on the Sersic index, i.e.,\,~the radial profile of the surface brightness, following equation (20) of \citet[][]{Cappellari2006} \citep[cf.][]{Bertin2002}. $\sigma_{R}$ denotes the stellar velocity dispersion within an aperture of radius $R$ and $G$ is the gravitational constant. We used the Sersic indices measured from the \textit{i}-band NGVS images and derived the velocity dispersion within the largest ellipse covered by our observations to estimate $\sigma_{Re}$ using the equation (1) of \cite{Cappellari2006}. For an aperture of one effective radius, this mass is representative of the enclosed dynamical mass \citep{Cappellari2006, Cappellari2013XV}. Our derived masses are presented in Table~\ref{tab:dataresults}.\

\begin{deluxetable*}{ccccccccccc}
 \tablewidth{0pt}
 \tablecaption{Derived Kinematic Parameters, Ages and Abundances
 \label{tab:dataresults}}
 \tablehead{
  \colhead{Galaxy} &
  \colhead{$R_{e}$} &
  \colhead{n} &
  \colhead{$\lambda_{R}$} &
  \colhead{$\sigma_{R}$} &
  \colhead{$\sigma_{R_{e}}$} &
  \colhead{M$_{e}$} &
  \colhead{Age (SSP)} &
  \colhead{[Z/H] (SSP)} &
  \colhead{${\Sigma_{15}}$} &
  \colhead{${L_{max}/L}$} \\
  \colhead{} &
  \colhead{(pc)} &
  \colhead{} &
  \colhead{} &
  \colhead{(\kms)} &
  \colhead{(\kms)} &
  \colhead{(\Msun)} &
  \colhead{(Gyr)} &
  \colhead{(dex)} &
  \colhead{(gal.deg$^{2}$)} &
  \colhead{} \\
  \colhead{(1)} &
  \colhead{(2)} &
  \colhead{(3)} &
  \colhead{(4)} &
  \colhead{(5)} &
  \colhead{(6)} &
  \colhead{(7)} &
  \colhead{(8)} &
  \colhead{(9)} &
  \colhead{(10)} &
  \colhead{(11)} } 
  \startdata
 	VCC\,0032 & 447.9 & 2.39 & 0.0915$^{1}$ & 30.9  & 29.5$\pm0.72$  & (4.5$\pm0.22$) $\times\,10^{8}$  &  3.5$_{-0.9}^{+3.2}$  &  -0.687$_{-0.053}^{+0.282}$ & 13.43  & 3.04 \\ 
	VCC\,1178 & 486.0 & 2.52 & 0.3959$^{1}$ & 138.3 & 132.1$\pm3.21$ & (9.6$\pm0.46$) $\times\,10^{9}$  &  9.1$_{-2.3}^{+2.9}$  &  -0.052$_{-0.137}^{+0.091}$ & 82.99  & 61.89 \\   
	VCC\,1192 & 408.9 & 1.81 & 0.3358$^{1}$ & 56.3  & 53.7$\pm1.31$  & (1.4$\pm0.07$) $\times\,10^{9}$  &  10.4$_{-4.5}^{+2.2}$ &   0.038$_{-0.173}^{+0.176}$ & 98.54  & 283.1\\  
	VCC\,1199 & 165.3 & 1.19 & 0.2657$^{1}$ & 53.4  & 51.1$\pm1.24$  & (5.5$\pm0.27$) $\times\,10^{8}$  &  9.1$_{-3.6}^{+3.4}$  &   0.22$_{-0.107}^{+0.0}$  & 93.41  & 968.2\\ 
	VCC\,1297 & 208.6 & 2.45 & 0.1599$^{1}$ & 228.1 & 217.9$\pm5.30$ & (1.1$\pm0.05$) $\times\,10^{10}$ &  11.2$_{-1.8}^{+1.3}$ &   0.22$_{-0.182}^{+0.0}$  & 133.50 & 122.35\\
	VCC\,1440 & 629.1 & 3.14 & 0.1672$^{2}$ & 34.8  & 32.4$\pm1.25$  & (7.0$\pm0.53$) $\times\,10^{8}$  &  8.7$_{-3.8}^{+3.8}$  &  -0.415$_{-0.127}^{+0.096}$ & 30.29  & 3.24 \\
	VCC\,1475 & 739.5 & 3.32 & 0.0657$^{2}$ & 82.8  & 77.0$\pm2.96$  & (4.5$\pm0.35$) $\times\,10^{9}$  &  12.5$_{-0.1}^{+0.0}$ &  -0.415$_{-0.057}^{+0.055}$ & 25.64  & 0.11 \\
	VCC\,1627 & 280.7 & 1.62 & 0.3155$^{1}$ & 40.5  & 38.7$\pm0.94$  & (5.2$\pm0.25$) $\times\,10^{8}$  &  6.5$_{-3.7}^{+3.1}$  &   0.038$_{-0.228}^{+0.181}$ & 99.28  & 85.82			
\enddata

\tablecomments{(1) Galaxy name from \citet{Binggeli1985} catalogue, (2) Effective radius in pc derived from the effective radius and distance indicated in Table~\ref{tab:data} (columns 10 and 4, respectively), (3) Sérsic index from NGVS \textit{i}-band images, (4)-(5) Specific stellar angular momentum $\lambda_{R}$ and velocity dispersion within an aperture of ${R_{e}/2}^{1}$ or ${R_{e}/3}^{2}$, (6)-(7) Estimated velocity dispersion and dynamical masses at at one effective radius using \citet{Cappellari2006}, (8)-(9) SSP-equivalent age and metallicity measured within an aperture of ${R_{e}/2}^{1}$ or ${R_{e}/3}^{2}$. The errors on our measurements are taken to be where the marginalized $\chi^{2}$ surface of the fit shows a difference of 1 with respect to the minimum value in the log(age)-[Z/H] parameter space. (10) Galaxy surface density $\Sigma_{15}$, i.e.\,, within a radius including the 15 brightest neighbors, derived from NGVS data, (11) Luminosity ratio of the brightest galaxy among the 15 neighbors of the galaxy and the galaxy.}
\end{deluxetable*}

We then combined our sample with 39 (on average) more extended objects from \cite{Toloba2014bStellPop} observed with long-slit part of the SMACKED survey, and representative of the low-mass ETGs of the Virgo cluster within the magnitude range $-19.0< M_{r}<-16.0$. Dynamical masses from these objects were derived by \cite{Toloba2014bStellPop} in a similar way to ours, using the S\'{e}rsic-dependent virial estimate. We also compared the NGVS effective radius values with those of \cite{Toloba2014bStellPop} for the objects in common and found a very good agreement. Therefore we can be confident in the relative position of our sample compare to \cite{Toloba2014bStellPop} sample in the mass-size diagram. We also extended the analysis to significantly higher masses using 56 Virgo ETGs observed as part of the \atlas\ project. The sizes and masses for the \atlas\ sample come from \cite{Cappellari2013XV, Cappellari2013XX}. The masses were derived via anisotropic Jeans dynamical modeling, which were shown to give good agreement with S\'{e}rsic-dependent virial mass estimates \citep{Cappellari2013XV}. As expected, Fig.~\ref{fig:sizemassDensity} (panel \textit{a}) shows that our sample contains low-mass ETGs that are more compact than those targeted in \citealt{Toloba2014bStellPop}. Indeed, for the same dynamical mass ($10^{8}$--$10^{10}$\,\Msun\,) our galaxies have effective radii that are nearly an order of magnitude smaller than their extended counterparts: {\em thus, our study probes a new parameter space for low-mass ETGs.} Clearly, it will be of interest in future IFU surveys to target galaxies in this under-represented part of parameter space. Note that the prominent gap in mass between \cite{Toloba2014bStellPop} and \atlas\, ETGs in Fig.~\ref{fig:sizemassDensity} is purely due to sample selection effects. Two objects from our sample, namely VCC\,1178 and VCC\,1297, are of particular interest as they overlap in mass with the lower-mass range of the \atlas\ objects (see also Fig.~\ref{fig:sizemassA3D}), but are located within the `zone of exclusion' (ZOE) defined by \citep{Cappellari2013XX}. Both objects have an absolute K-band magnitude ($M_{K}$=-21.4 and -20.8, respectively) that is fainter than the \atlas\ sample selection definition ($M_{K} < -21.5$). At face value, this implies an unusually high mass-to-light ratio for these objects, but detailed modeling would be required to determine the cause of this, which is outside the scope of this paper.\

In Fig.~\ref{fig:sizemassDensity}, the galaxy density is expressed in two different flavors: (1) the luminosity of the brightest galaxy among its 15 closest neighbors, normalized by the luminosity of the target galaxy, $L_{max} / L$; and (2) the surrounding galaxy density, $\Sigma_{15}$, defined as the number of galaxy per square degree within a projected region that includes the 15 closest neighbors to the target galaxy. These two values for our sample galaxies are presented in Table~\ref{tab:dataresults}. The first result that emerges from these plots (Fig.~\ref{fig:sizemassDensity}) is that the five low-mass galaxies that depart most dramatically from the main mass-size sequence (VCC\,1178, VCC\,1192, VCC\,1199, VCC\,1297, VCC\,1627) are all found in regions of high galaxy density (i.e.\,,~high $\Sigma_{15}$, see panel \textit{b}) and, most importantly, that also contains a massive galaxy (i.e.\,,~high $L_{max} / L$, see panel \textit{a}), as expected since they are in close proximity to either M\,87~(NGC\,4486), M\,49~(NGC\,4472) or M\,89~(NGC\,4552). The fact that most galaxies above $10^{10}$~M$_{\odot}$ have low $L_{max} / L$ parameters is a natural consequence of the galaxy luminosity function: i.e., the probability that such a galaxy will be found next to an even more massive galaxy is lower than for low-luminosity objects. In the low mass regime ($< 10^{10}$~M$_{\odot}$), we indeed see a much wider range of $L_{max} / L$ values, which makes the high $L_{max} / L$ value of the more compact low-luminosity galaxies even more relevant. Interestingly, these five objects have very small numbers or no globular cluster systems \citep{Peng2008}, suggesting indirectly that they could have experienced tidal stripping events. If, rather than using $L_{max} / L$, we use the normalized {\em total} luminosity $L_{tot} / L$, the results do not change.  

To better reveal the first order relations in the three-dimensional space, we also smoothed the data using the two-dimensional Locally Weighted Regression (LOESS) method (Cleveland \& Devlin 1988), implemented as described in \cite{Cappellari2013XX}. Results are shown in the two bottom panels of Fig.~\ref{fig:sizemassDensity} (panels \textit{c} and \textit{d}). This analysis used a regularization factor of 0.3.\

Ultimately, we performed the same exercise with a various number of closest neighbors N (i.e.\,~N=3, 10 and 25) and we did not see any significant changes in the trends except a systematic increase of $L_{max} / L$ with N (as expected from the cluster luminosity function).

\subsection{Rotational support: the $\lambda_{R}-\epsilon$ diagnostic}
\label{sec:LR}

\begin{figure}
	\includegraphics[width=\columnwidth]{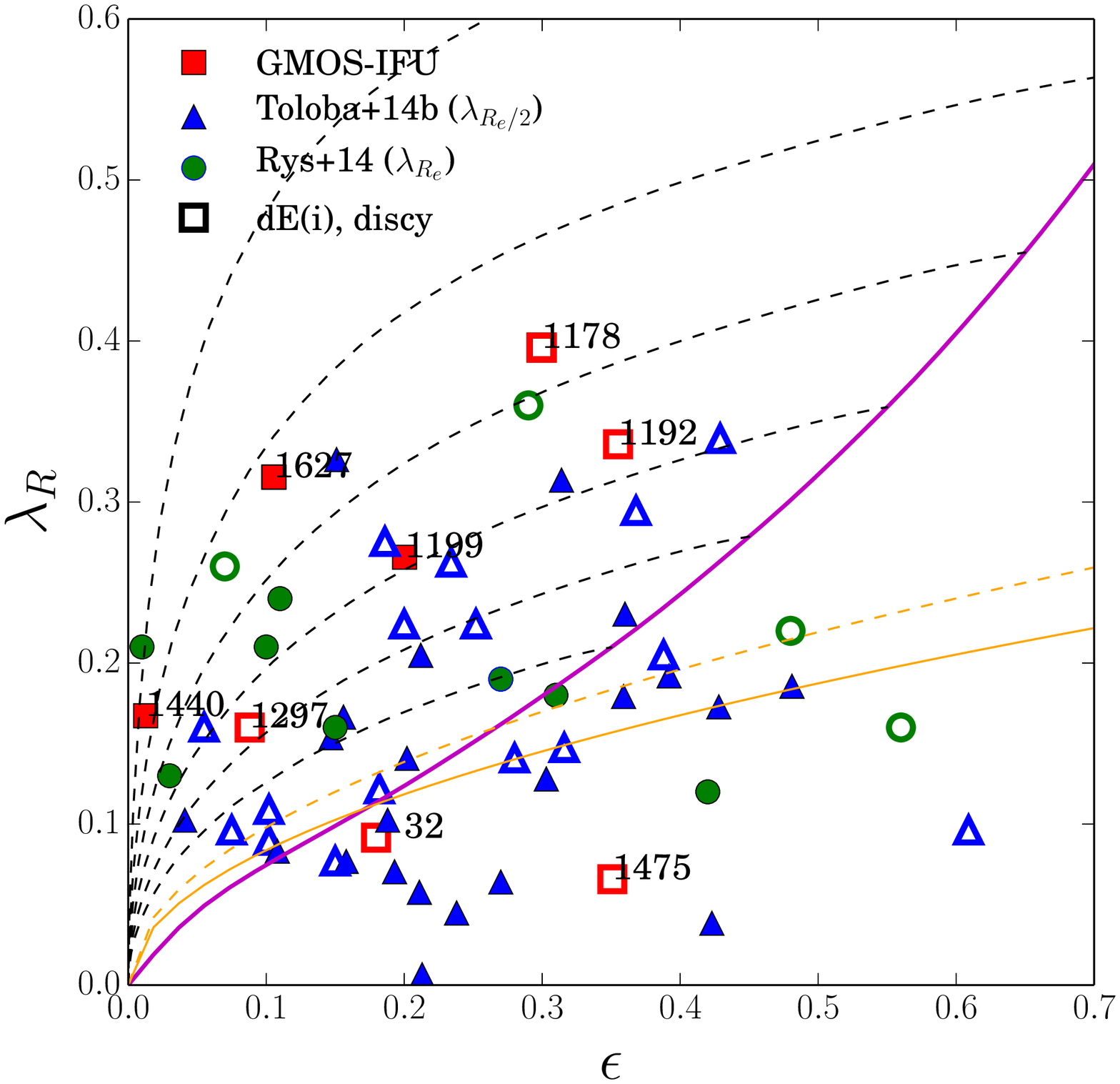}
	\caption{${\lambda_{R}-\epsilon}$ diagram. Values have been derived at $R_{e}/2$ for all our \gmosifu\ objects except VCC\,1440 and VCC\,1475, derived at $R_{e}/3$. ${\lambda_{Re}}$ values of more extended low-mass ETGs in the Virgo cluster from \citet{Rys2014} are shown as green circles and ${\lambda_{Re/2}}$ values from \cite{Toloba2014cLR} as blue triangles. For \cite{Toloba2014cLR} and \cite{Rys2014} samples, open symbols represent objects with underlying disky-structure identified in \cite{Lisker2007a}. For \gmosifu\, sample, open symbols represent objects with disky isophotes identified in the NGVS. The magenta line represents the empirical relation between $\lambda_{R}$ and $\epsilon$ for an ellipsoidal object seen edge-on and having an anisotropy parameter of $\beta=0.65\times\epsilon$. The black dashed lines correspond to this same relationship for an object seen edge-on to face-on. The orange line (respectively the dashed orange line) represents the empirical relation $\lambda_{R}=0.265\times\sqrt{\epsilon_{e/2}}$ (respectively $\lambda_{R}=0.31\times\sqrt{\epsilon_{e}}$) defined by \citet{Emsellem2011FRSR} which separates slow-rotators from fast-rotators.}
	\label{fig:LR}
\end{figure}

\citet{Emsellem2011FRSR} used the specific stellar angular momentum versus ellipticity diagram (${\lambda_{R}-\epsilon}$) to classify massive ETGs, reporting an empirical criterion with which to identify slow and fast rotators (that represent $\sim$~15\% and 85\% of the \atlas\ sample, respectively). This scheme has been used subsequently to constrain the properties of possible low-mass ETGs progenitors \citep{Rys2013, Toloba2014cLR}. Accordingly, we followed the methodology given in \cite{Emsellem2007LR} to derive ${\lambda_{R}}$ profiles for all galaxies in our sample (up to the maximum radius coverage of the \gmosifu\ FOV). Our FOV limit corresponds to $R_{e}/2$ for all galaxies in our sample except VCC\,1440 and VCC\,1475, which are only covered up to $R_{e}/3$. The measured values of ${\epsilon}$ are presented in Table~\ref{tab:data}, and \lr\, in Table.~\ref{tab:dataresults}. The ${\lambda_{R}-\epsilon}$ diagram is shown in Fig.~\ref{fig:LR}.

We find six fast-rotators (75\%) among our sample of eight galaxies: VCC\,1178, VCC\,1192, VCC\,1199, VCC\,1297, VCC\,1440 and VCC\,1627. The two remaining galaxies are classified as slow-rotators: VCC\,0032 and VCC\,1475. The fraction of slow-rotators in low-mass compact ETGs (25\%) is slightly higher than that found among high-mass ETGs, but given the small number of objects in our sample, the difference is not significant. Still, it is worth pointing out that the fraction of slow-rotators among the entire low-mass ETGs population (i.e., the \citealp{Rys2014, Toloba2014cLR} and present \gmosifu\, sample combined) is still at least 25\%, or 10\% higher than in high-mass ETGs. Interestingly, VCC\,0032 shows gradual but large variations in both ellipticity and position angle within its inner 40\arcsec, as well as disky-isophotes. It may therefore represent a true slow rotator, as opposed to a more face-on fast rotator. The low ${\lambda_{R}}$ value of VCC\,1475 is not unexpected since we have shown it to be an unambiguous ${2-\sigma}$ galaxy \citep[][see \S~\ref{sec:VCC1475}]{Krajnovic2011}. 

Our sample galaxies are found to have $0.1 \lesssim {\lambda_{R_{e}/2}} \lesssim 0.4$ and relatively low ellipticities: $\epsilon \le 0.4$. This is similar to what has been found for more extended low-mass ETGs by both \citet{Rys2014} and \citet{Toloba2014cLR}. Note that the \cite{Rys2014} \lr\ values are derived at ${R_{e}}$, creating a systematic, but not significant, positive offset. None of our targets reach the high values of ${\lambda_{R_{e}/2}}$ $> 0.4$ or $\epsilon > 0.4$ that are commonly observed in higher mass ETG samples \citep{Emsellem2011FRSR}. 

\begin{figure*}
	\includegraphics[width=\hsize]{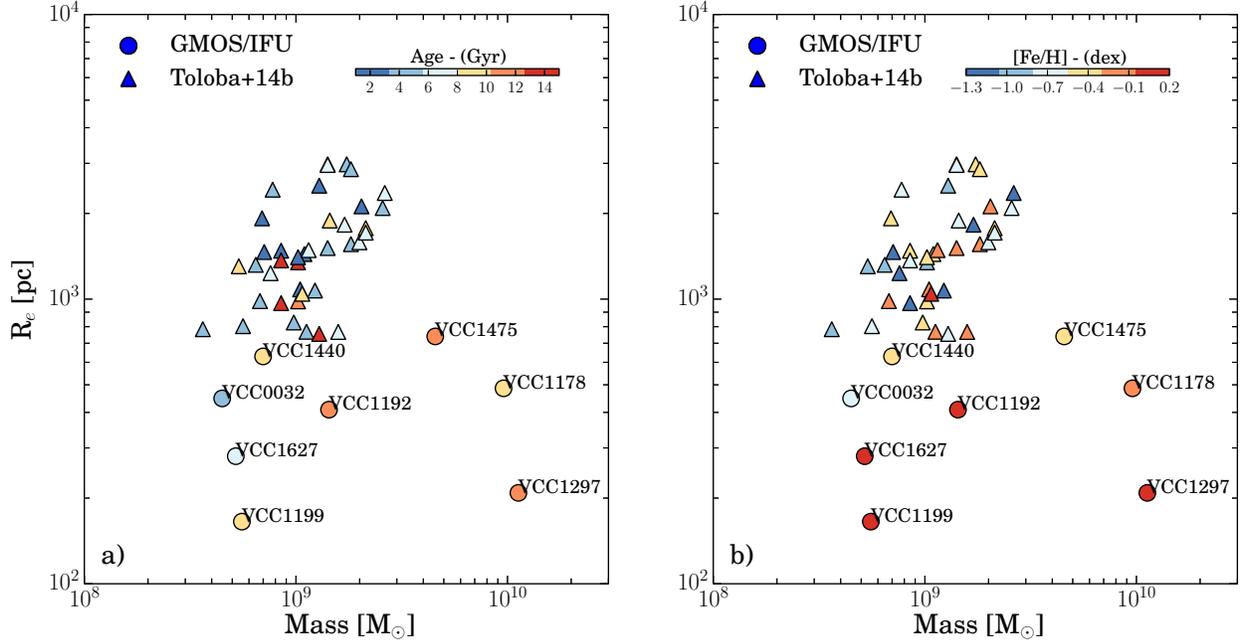}
    \caption{Location of our \gmosifu\ galaxies (circle symbols) in the mass-size diagram combined with those of \citet{Toloba2014bStellPop} (triangles) observed with long-slit. The color of each symbol corresponds to the SSP-equivalent age (panel \textit{a}) and metallicity (panel \textit{b}) of the combined spectra within $R_{e}/2$, as indicated by the color bar in each panel.}
	\label{fig:sizemass}
\end{figure*}

\cite{Toloba2014cLR} suggested the existence of a correlation between ${\lambda_{R}}$ and the projected distance to the center of the cluster (e.g., M\,87) as well as the presence of subtle disky structures in fast-rotators. Among our targets, VCC\,0032, VCC\,1475 and VCC\,1440, are the outermost galaxies. They are the most extended objects and are located in regions of relatively low galaxy density. The five remaining objects (VCC\,1178, VCC\,1192, VCC\,1199, VCC\,1297 and VCC\,1627) are found close to, either M\,87 or a large companion (see \S\ref{sec:galdensity}) and are all fast-rotators. More interestingly, they are faster rotators than the three remaining objects and are also the most compact objects of our sample. While these results may appear in contradiction with the trend suggested by \cite{Toloba2014cLR} --- that, on average, $\lambda_{R_{e}}$ normalized by $\sqrt{\epsilon}$ \citep[see][]{Emsellem2011FRSR,Toloba2014cLR} is lower at small distances from M\,87 --- their trend seem weaker, or even absent (see their Figure~5), for galaxies with disk-like structures. Three of the five compact galaxies may have such disk features, suggested by their disky isophotes, so it may not be surprising if they do not participate in such a trend. Observations of larger galaxy samples are definitively needed to confirm, or not, these possible trends.

\subsection{Stellar content and star formation history}
\label{discussion_stellpop}

\begin{figure*}
	\includegraphics[width=\hsize]{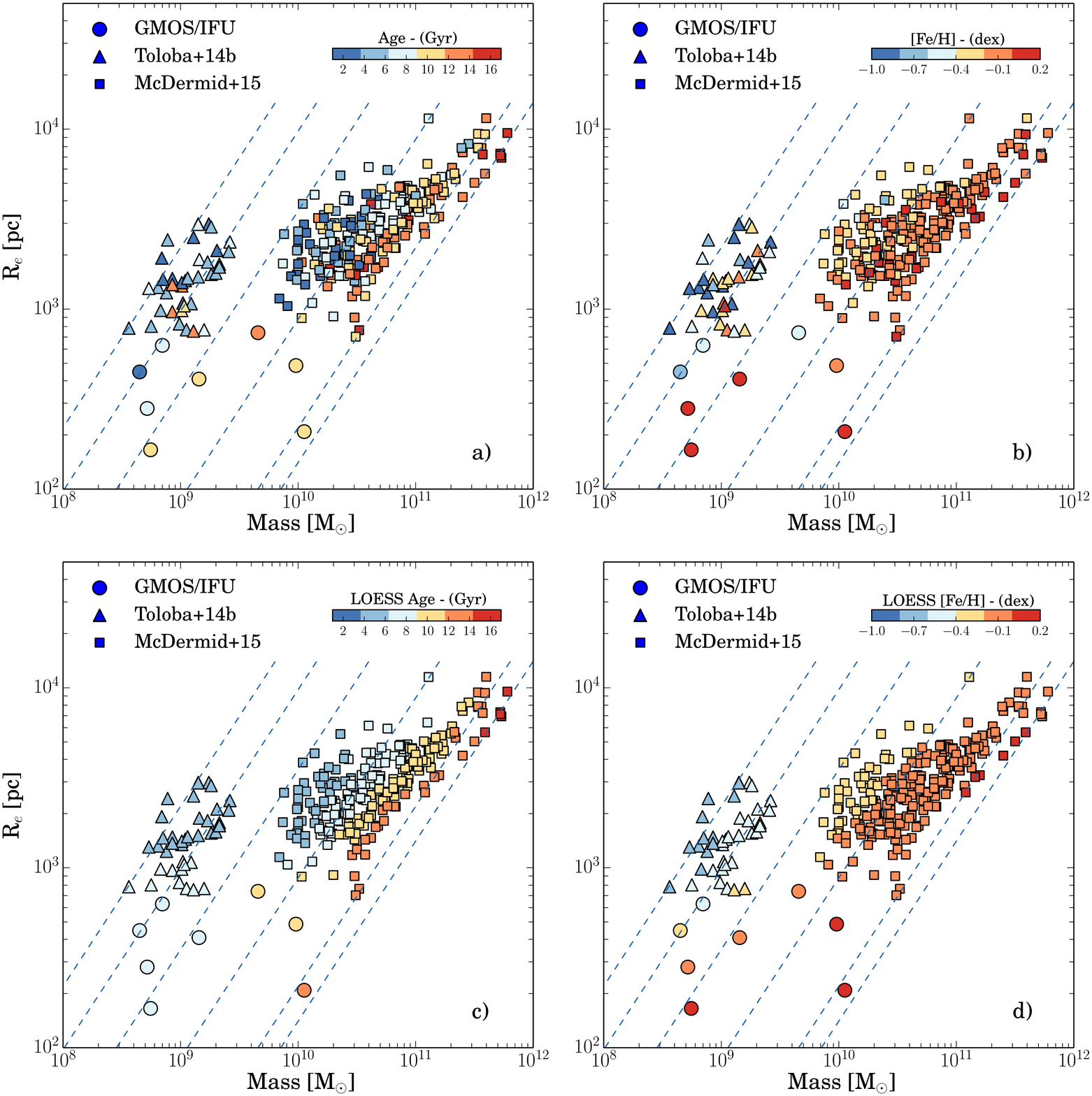}
    \caption{Location of our \gmosifu\ galaxies (circle symbols) in the mass-size diagram. Samples from \citet{Toloba2014bStellPop} observed with long-slit (triangle symbols), and \citet{McDermid2015} from the \atlas\, survey (square symbols), had been added to the diagram. The colors of each symbol correspond to the SSP-equivalent age (left panel) and metallicity (right panel) as indicated by the color bars in each panel, respectively. The light-blue dotted lines represent the lines of constant velocity dispersion ${\sigma}$=20, 30, 50, 100, 200 \& 250 \kms\,(from left to right) as implied by the virial mass estimator ${M=5R_{e}\sigma^{2}/G}$. The gap between the low- and high-mass ETGs is due to sample selection effects.}
	\label{fig:sizemassA3D}
\end{figure*}

We now analyze the age and metallicity of the stellar populations in our target galaxies, as tracers of their formation and assembly history. Because the SSP-equivalent ages and metallicities were derived from the stacked spectra, they reflect the single stellar population that dominates the light within the central regions, i.e\,~$R_{e}/2$ or $R_{e}/3$ (see Table~\ref{tab:dataresults}). No trend is found between the stellar population parameters and the location of our targets within the cluster (the local galaxy density) or their $\lambda_{R}$ parameter. This is qualitatively in agreement with \cite{Toloba2014bStellPop}. It is, however, interesting to note that all objects with underlying disky isophotes, aside from VCC\,0032, show the oldest single stellar population.

As emphasized in \S\ref{sec:galdensity}, the galaxies in our sample are more compact than those examined in other studies of low-mass ETGs. Thus, we investigated the potential link between their compactness and their stellar population age and metallicity as shown in Fig.~\ref{fig:sizemass}. Here we present the \gmosifu\ sample in terms of their SSP-equivalent age and metallicity distributions in the mass-size diagram. To compare our sample to more extended low-mass ETGs, we have added the objects from \cite{Toloba2014bStellPop}, shown as triangles. No clear trend appears between the age and the location of the objects in the mass-size plane although, on average, the compact ETGs are older than less compact systems (panel \textit{a} of Fig.~\ref{fig:sizemass}). A possible trend is visible between the metallicity and the compactness of low-mass ETGs (panel \textit{b} of Fig.~\ref{fig:sizemass}): e.g., more compact objects tend to be more metal-rich. In the context of low-mass ETGs being formed through stripping and harassment, \cite{Rys2014}, showed that these objects become more compact than their progenitors, which are often assumed to be late-type, low-mass spiral galaxies. The \gmosifu\ sample includes more compact objects than those in \cite{Toloba2014cLR} and \cite{Rys2014}, so we would expect our objects to have been further processed. Our findings of older and more metal-rich stellar populations in the most compact ETGs is fully consistent with this picture.

In the broader context of galaxy formation, there exists a longstanding debate on whether low- and high-mass ETGs show a dichotomy in their properties, i.e.\,, whether they represent two distinct populations (as first advocated by \citealt{Kormendy1985}), or whether they form a more continuous sequence in terms of their structure, kinematics and stellar populations. Most recent studies of photometric scaling relations and isophotal parameters suggest a continuity between the low- and high-mass regimes \citep{Jerjen1997, Graham2003, Gavazzi2005, Ferrarese2006, Cote2007}, but an alternative view is favoured by \citet{Kormendy2009}. Recent surveys, especially those using IFU spectroscopy, have characterized the detailed kinematics and stellar population properties of more massive ETGs. Using them as a reference sample, we have combined the 260 massive ETGs ($10^{10}$\Msun~$<M<10^{12}$~\Msun) from  \atlas\  \citep{Cappellari2013XV, McDermid2015}, with the 39 low-mass ETGs from \cite{Toloba2014bStellPop}  and our eight compact low-mass ETGs. We focus here on the SSP-equivalent stellar population information (e.g.\,~age and metallicity) as projected onto the mass-size diagram (see Fig.~\ref{fig:sizemassA3D}). As in Fig.~\ref{fig:sizemassDensity}, we smoothed the raw data (panels \textit{a} and \textit{b} of Fig.~\ref{fig:sizemassA3D}) using the LOESS method with a regularization factor of 0.3 to reveal the first order relationship in the three-dimensional space (panels \textit{c} and \textit{d} of Fig.~\ref{fig:sizemassA3D}). 

The first striking result from this figure is that there is a level of continuity in age and metallicity for ETGs between $10^{8}$~\Msun\, and $10^{12}$~\Msun\,. As observed in \cite{McDermid2015}, at fixed size, more massive galaxies are older; alternatively, at fixed mass, larger galaxies are younger. In term of metallicity, low-mass ETGs that are more compact are also more metal-rich, and more massive galaxies in general tend to have more metal-rich stellar population \citep{Thomas2010, McDermid2015}. It is also interesting to note that the ages and metallicities of the most compact, low-mass ETGs are more representative of more massive ETGs, rather than of less compact galaxies with comparable mass. This similarity between  compact low-mass and massive ETGs has already been observed, in term of metallicity only, by \citealt{Chilingarian2009sci} for a sample of 21 compact ellipticals in the local group. This is in agreement with \cite{McDermid2015} who showed that the stellar population parameters of massive ETGs in the mass-size plane roughly follow lines of iso-velocity dispersion~(see also \citealt{Cappellari2011a}), as indicated by the light-blue dotted lines in Fig.~\ref{fig:sizemassA3D}. In fact, we find that this result can be extended to low-mass ETGs as shown in Fig.~\ref{fig:sizemassA3D} (albeit with a large scatter in the age distribution), with the transition between low- and high-mass ETGs being smooth. Taken at face value, this naively suggests that low-mass ETGs ($10^{8}<M<10^{10}$~\Msun) may have been shaped by the same secular and/or environmental mechanisms as their massive counterparts ($M>10^{10}$~\Msun). However, it does not necessarily imply that galaxies evolve along iso-velocity dispersion lines, neither constant metallicity lines. Only detailed simulations focused on stellar population and kinematic parameters would give us clues about the evolutionary track in each scenario (e.g.\, tidal stripping, ram pressure stripping, harassment, etc.). Fig.~\ref{fig:sizemassA3D} mostly confirms that velocity dispersion is a good predictor of the age and metallicity of galaxies \citep{Graves2010, Wake2012, McDermid2015}, even for low-mass ETGs. 

Finally, if we consider the two-dimensional, mass-weighted age and metallicity maps for our objects (see Fig.~\ref{fig:maps} and \S~\ref{sec:stellarpop}), we see that nearly half our program objects contain a younger and more metal-rich stellar population in their cores. This result is in agreement with what has been previously observed in ``dwarf" galaxies \cite[e.g.\,,][]{Lisker2006b, Michielsen2008, Chilingarian2009a, Spolaor2010a, Rys2013}, as well as in many more massive ETGs by \cite{kuntschner2010}. This finding adds further support to the notion of a continuity among low-mass and high-mass ETGs.

\subsection{Formation scenarios}
\label{sec:scenarios}
Over the past decade, many studies have argued for a scenario in which the low-mass (``dwarf") ETGs in clusters originated from late-type, star-forming galaxies that were transformed by the cluster environment \citep{Moore1998, Boselli2008a, Rys2014, Toloba2009, Toloba2014cLR}. Two main mechanisms are usually invoked to transform the infalling objects. First, ram-pressure stripping removes gas in about 100--200~Myr and rapidly quenches the star formation on a time-scale of $\sim~1$~Gyr \citep{Boselli2008a}. Second, gravitational tidal harassment dynamically heats the low-mass ETG progenitors, lowering their angular momentum and concentrating the stellar component (leading to more compact systems; \citealt{Rys2014}). A third process is certainly also at play: starvation \citep{Larson1980}, which is a form of gas stripping but acts on a much longer time scale ($>$5~Gyr) as shown by \citealt{Boselli2009} (see their Fig.~4), and affects more efficiently the gaseous halo rather than the inner regions. This process tends to significantly decrease the surface brightness of the affected object but keeps its characteristic size rather unchanged, in contrast with the effect of ram-pressure stripping~\citep[see Fig.~3 of][]{Boselli2009}. Starvation is therefore not able to solely explain the properties of low-mass ETGs, especially for the more compact ones. However it certainly helps stopping their star formation, especially in clusters where a high rate of interactions can easily remove the gas~\citep{Rasmussen2012}.

The exact nature of the progenitors is still a matter of debate. \cite{Lisker2013}, based on a comparison between Virgo observations and Millenium II simulations,  pointed out that late-type galaxies in the local universe may not be representative of the progenitors of today's low-mass ETGs. They computed the time spent by a low-mass ETG inside a high-mass ($M>10^{12}$~\Msun) halo, as well as the mass-loss since first entering the halo (see their Fig.~4). According to their analysis, and using the galaxy density parameter $\Sigma_{10}$ derived from the NGVS (see \S~\ref{sec:galdensity}), our sample galaxies would have spent, on average, $\sim$ 7~Gyr inside a halo more massive than $10^{12}$~\Msun\,. Since our objects have a median age of 9~Gyr, this suggests that they have spent most of their lives inside a massive halo, experiencing a variety of environmental effects (i.e.\,~from close pair to group interactions) that would certainly have helped shape their present day structure. Because the scatter in these correlations is quite large, it is hardly surprising for some galaxies to undergo infall at later times and/or experience quite different interaction histories. This stochasticity could explain the diversity in kinematics and stellar populations observed in low-mass ETGs, as well as the highly compact nature of some (rare) objects in dense regions.

The galaxies in our sample have $\lambda_{R}$ values that are lower than more massive galaxies but consistent with ``typical" values found for low-mass ETGs. Gravitational harassment (including tidal stripping) could explain such a decrease in angular momentum and one might naively expect that the longer a given galaxy stays in the cluster, the more angular momentum it looses and the more compact it becomes. However, our galaxy sample does not show a systematically lower specific angular momentum than less compact, low-mass ETGs. This may suggest either that their compactness is not related solely to gravitational harassment, or that this process (especially tidal stripping) can sometimes raise the stellar angular momentum of the system.

Several of our program galaxies harbor younger and more metal-rich stellar populations in their center compared to their outskirts. This is consistent with what one would expect from ram-pressure stripping, where the pressure of the intra-cluster medium can remove gas from a galaxy but its efficiency depends on the relative strength of the gravitational potential of the object. The gas is thus more easily removed from the outskirts of a galaxy than from its center where it can stay bound to its host and continue forming stars \citep[see][in the case of a late-type galaxy]{Fumagalli2014}. Gravitational tidal forces could also be responsible for younger stellar populations in the centers of low-mass ETGs as it promotes the formation of bars \citep{Mayer2006} that are known to induce gas infall. Others scenarios could potentially explain these observations, though: simple secular evolution, external accretion of gas, i.e.\,~from the intergalactic medium or from a gas-rich companion (interaction, merger). The latter is more likely to occur in small groups, before a galaxy enters the dense regions of the present-day cluster where the relative velocities are high.

The existence of ${2-\sigma}$ galaxies in the low-mass ETGs population may also provide some constraints pertaining to their formation and evolution. The counter-rotating disk in VCC\,1475, as revealed by our two-dimensional spectroscopy, appears as a rather thin, central structure. N-body simulations predict that $\sim1$\% of the dwarfs in Virgo could exhibit counter rotation due to harassment \citep{Gonzalez2005}. However, the formation of such a counter-rotating structure via this process would make the outer part of the galaxy ($R>R_{e}$) counter rotating, and this is not what is observed for VCC\,1475. A pure stellar merger event is also unlikely in this case as the inner disk is more metal-rich than the surrounding stellar component. A gas-accretion event is a more promising hypothesis, probably the result of an interaction with a gas-rich galaxy \citep[see][for more details about KDC formation scenarios in cluster environments]{Toloba2014a}. The exact origin of this $2-\sigma$ galaxy is unclear but its appearance in a low-mass ETG sample certainly adds a level of complexity to the stripping and harassment scenario, emphasizing the importance of the pre-cluster environmental history of cluster galaxies.

In summary, we have shown that the properties of the stellar populations in low-mass ETGs show a level of continuity with more massive ETGs drawn from the \atlas\ survey (which itself is representative of the local volume within 40~Mpc). The age and metallicity follow the lines of iso-velocity dispersion (with a large scatter) suggesting that the formation and evolution processes that makes low-mass ETGs (whether compact or not) are not very different from the ones acting on high-mass ETGs, or at least that their impact vary smoothly with the mass, size and environment of the galaxy.

\section{Summary}
\label{sec:summary}

We have presented and analyzed spectroscopic observations for eight low-mass compact ETGs in the Virgo cluster using the \gmosifu\ instrument mounted on the Gemini North 8-m telescope. These objects were selected from \cite{Binggeli1985} and our own Next Generation Virgo Survey \citep[NGVS;][]{Ferrarese2012a} to probe a regime of ETG size and surface brightness that has largely been unexplored with IFU instruments. Our goal has been to characterize their basic kinematics and stellar populations in order to examine possible connections between this rare class of galaxy with more massive ($M>10^{10}$~\Msun) and less compact low-mass ETGs. We performed kinematics and stellar populations analysis using full spectral fitting. Our main results can be summarized as follows:

\begin{itemize}
\item We calculated the dynamical mass inside one effective radius for each of our objects and confirmed that these galaxies are more compact than most ETGs of comparable mass \citep{Toloba2014cLR,Rys2014}. The sample galaxies span a dynamical mass range of $4.5\times10^{8}$ to $1.1\times10^{10}$\,\Msun\, and have effective radii of $165$ to $740$~pc. Using the parameters $L_{max}/L_{gal}$ and $\Sigma_{15}$ as proxies of the local galaxy density, we have found that the most compact low-mass ETGs are preferentially found in high density regions and often close to a very massive galaxy. 

\item The two-dimensional maps of stellar radial velocity, \textit{V}, and velocity dispersion, \textit{$\sigma$}, of our objects show a large variety of kinematic structures: from non-rotating systems to those with well organized and high rotation, as well as velocity dispersion profiles that vary from flat to centrally peaked. Our data also provide unambiguous evidence for the $2-\sigma$ nature of the low-mass galaxy VCC\,1475. In most cases, low-mass compact ETGs with organized and high radial stellar velocity exhibit disky isophotes as revealed by the analysis of the NGVS deep images. 

\item We assessed the rotational support in these galaxies by calculating their specific apparent angular momentum, $\lambda_{R_{e}/2}$,  finding two slow-rotators and six fast-rotators. Our sample galaxies have $\lambda_{R_{e}/2}$ values similar to those found previously in more extended low-mass ETGs \citep{Rys2013, Toloba2014cLR} and none of them reach values higher than 0.4 as is sometimes observed in more massive galaxies \citep{Emsellem2011FRSR}. They are generally round objects and we do not find very flattened galaxies ($\epsilon>$~0.4). Three  of the five most compact low-mass ETGs have disky isophotes, and all of them are fast rotators with the highest specific angular momentum of our sample.

\item We found that our program galaxies contain old stars, between 6 and 11.5~Gyr, with metallicity between $-0.7$ and $0.2$. Half of our sample galaxies contain younger and more metal-rich stellar populations in their centers. Interestingly, low-mass compact ETGs with underlying disky isophotes are found to have the oldest stellar populations among our sample.

\item We derived  SSP equivalent ages and metallicities within $R_e$/2 for our program galaxies, finding a striking continuity between the low-mass and high-mass ETGs \citep{Cappellari2013XV}. The more compact galaxies, at a fixed mass, are found to have more metal-rich and older (albeit with a larger scatter) stellar populations. We also extended the results from \cite{McDermid2015} to low-mass ETGs, showing that the velocity dispersion is a relatively good predictor of the stellar population properties.\

\end{itemize}\

Our results are consistent with a scenario in which low-mass, compact ETGs have spent a large fraction of their lives in the cluster environment and have been strongly shaped by environmental mechanisms such as ram-pressure stripping, tidal stripping, starvation and gravitational harassment. This picture is quantitatively supported by a comparison of our data to the study of \cite{Lisker2013}, who jointly analyzed Millenium II simulations and Virgo galaxy observations. Our study underscores the importance of environmental history for low-mass ETGs, such as close-pairs, groups, etc. before the pre-present day cluster epoch. Perhaps most significantly, our study adds to a growing body of evidence that a continuity exists in the structural, kinematic and stellar population properties of low- and high-mass ETGs, which in turn suggests that ETG formation and evolution processes vary smoothly with mass, size and environment across the full range in galaxy mass. A more detailed account of the relative weight of each evolutionary processes should wait for observations of larger galaxy samples and realistic numerical simulations.

\section{acknowledgements}
\label{sec:acknow}
This work is based on observations obtained at the Gemini Observatory processed using the Gemini IRAF package, which is operated by the Association of Universities for Research in Astronomy, Inc., under a cooperative agreement with the NSF on behalf of the Gemini partnership: the National Science Foundation (United States), the National Research Council (Canada), CONICYT (Chile), the Australian Research Council (Australia), Ministério da Ciência, Tecnologia e Inovação (Brazil) and Ministerio de Ciencia, Tecnología e Innovación Productiva (Argentina).

The authors thank Elisa Toloba for sharing her work ahead of publication and Mina Koleva for the use of her data for VCC\,1475.\ 

AG acknowledges the hospitality and support of the Macquarie University Astronomy and Astrophysics Center during the preparation of this work.\

EWP acknowledges support from the National Natural Science Foundation of China under Grant No. 11173003, and from the Strategic Priority Research Program,``The Emergence of Cosmological Structures'', of the Chinese Academy of Sciences, Grant No. XDB09000105.\

This work is supported in part by the French Agence Nationale de la Recherche (ANR) Grant Programme Blanc VIRAGE (ANR10-BLANC-0506-01), and by the Canadian Advanced Network for Astronomical Research (CANFAR) which has been made possible by funding from CANARIE under the Network-Enabled Platforms program. This research used the facilities of the Canadian Astronomy Data Center operated by the National Research Council of Canada with the support of the Canadian Space Agency. The NGVS team owes a debt of gratitude to the director and the staff of the Canada-France-Hawaii Telescope, whose dedication, ingenuity, and expertise have helped make the survey a reality.\

{\it Facilities}: Gemini, CFHT




\label{lastpage}

\end{document}